\documentclass[twocolumn,pre]{revtex4-1}
\usepackage{amsthm,setspace,afterpage}
\usepackage[usenames,dvipsnames]{xcolor}
\usepackage{amsmath,amssymb,graphicx,bm,subfigure}
\usepackage{hyperref}
\usepackage{float}
\usepackage{ulem}

\begin{document}
\title{Orientational Phase Transitions and the Assembly of Viral Capsids
}

\author{Sanjay Dharmavaram$^{1}$, Fangming Xie$^{2}$, William Klug$^{3}$, Joseph Rudnick${^1}$, Robijn Bruinsma${^1}$}

\affiliation{${1}$Department of Physics and Astronomy, 
  University of California, Los Angeles,  CA 90095, USA}
\affiliation{${3}$Department of Mechanical and Aerospace Engineering, 
  University of California, Los Angeles,  CA 90095, USA}
\affiliation{${2}$Department of Physics, University of Science and Technology of China, Hefei, Anhui, China}

\begin{abstract}
We present a generalized Landau-Brazovskii free energy for the
  solidification of chiral molecules on a spherical surface in the
  context of the assembly of viral shells. We encounter two types of
  icosahedral solidification transitions. The first type is a conventional
  first-order phase transition from the uniform to the icosahedral state.
It can be described by a single icosahedral spherical harmonic of even
  $l$. The chiral pseudo-scalar term in the free energy creates secondary terms with chiral character but it does not affect the thermodynamics of the transition. The second type, associated
with icosahedral spherical harmonics with odd $l$, is anomalous. Pure odd $l$ icosahedral states are unstable but
stability is recovered if admixture with the neighboring $l+1$ icosahedral spherical harmonic is
  included, generated by the non-linear terms. This is in conflict with the principle of Landau theory
  that symmetry-breaking transitions are characterized by only a \textit{single} irreducible
  representation of the symmetry group of the uniform phase and we argue that this principle 
should be removed from Landau theory.  The chiral term now directly affects the transition because it lifts the degeneracy between two isomeric mixed-$l$
icosahedral states. A direct transition is possible only
  over a limited range of parameters. Outside this range, non-icosahedral
  states intervene. For the important case of capsid assembly dominated by $l=15$, the intervening states are found to be based on octahedral symmetry.
\end{abstract}

\maketitle
\section{Introduction}
The seminal work of Onsager on nematic liquid crystals in the 1940's \cite{Onsager} initiated the theoretical study of \textit{orientational phase transitions}. Orientational phase transitions have since been investigated not only for liquid crystals but also for quasi-crystals, supercooled liquids, metallic glasses, atomic clusters and Fermi liquids \cite{Kats,Steinhardt}. Theories of orientational phase transitions are usually expressed in terms of the orientational order parameter $Q_{lm}$ defined as the coefficient of the spherical harmonic $Y_l^m$ in an expansion of the density $\rho(\Omega)$ of molecules oriented along the solid angle $\Omega$ \cite{Steinhardt}. In Landau theory, the order parameter associated with a continuous phase transition is characterized by a \textit{single} irreducible representations (``irrep") of the symmetry group $G_0$ of the disordered phase \cite{Kats}. If $G_0$ is the group $SO(3)$ of rotations in three dimensions then the order parameter should be characterized by just a single value for $l$. The remaining $2l+1$ coefficients $Q_{lm}$ of the order parameter must be determined by minimization of a Landau free energy functional expressed as an expansion of the $Q_{lm}$. The symmetry group $G$ of the ordered phase must be an isotropy subgroup of $G_0$.

The literature on orientational phase transitions has been mostly restricted to $l$ values less than or equal to $l=6$, but recently an application of orientational transitions involving larger values of $l$ has emerged in the area of \textit{viral assembly} ~\cite{lorman2008landau, lorman2007density}. Simple viruses are composed of a protein shell -- the capsid -- that surrounds the viral RNA or DNA genome molecule(s). Fig. 1 shows a cross-section of the reconstruction of a typical small, single-stranded (ss) RNA  virus -- the flock house virus (FHV) -- obtained by X-ray diffraction \cite{Johnson1994}.
\begin{figure}[htbp]
\begin{center}
\includegraphics[width=1.5in]{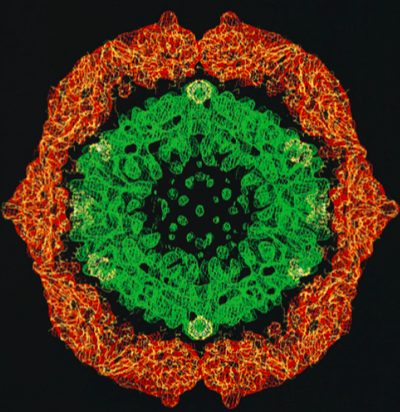}
\caption{Cross-section of a reconstruction of the flock house virus as obtained by x-ray diffraction viewed along a two-fold icosahedral axis (J.Johnson, private communication). Red: capsid composed of $180$ identical proteins. Green: enclosed, single-stranded RNA genome molecule (online in color). The diameter is 35 nanometers. }
\end{center}
\label{FHV1}
\end{figure}
The capsids of most spheroidal viruses, including FHV, have \textit{icosahedral symmetry}
\cite{caspar, baker}, \footnote{Spheroidal retroviruses, like the Rous Sarcoma Virus \cite{Kingston}, and spheroidal bunyaviruses, like the Uukuniemi virus, are examples of spheroidal capsids that lack icosahedral symmetry \cite{Overby}.}  Small RNA viruses like FHV have diameters in the 20-30 nanometer (nm) range and are composed of hundreds of nanometer-sized capsid proteins, which in the simplest case are all identical.  The assembly of small single-stranded (ss) RNA viruses, such as FHV, is driven by affinity between different capsid proteins (primarily hydrophobic attraction) and by affinity between the ss RNA genome molecules and the capsid proteins. The RNA-protein affinity has in general both a non-specific, electrostatic component and a specific component with binding of particular RNA motifs such as ``hairpins" to capsid proteins. Following the early work of Fraenkel-Conrat and Williams \cite{Fraenkel}, it was found that many small RNA viruses can assemble spontaneously under \textit{in-vitro} conditions in solutions containing the capsid proteins and the RNA genome molecules. The self-assembly of \textit{empty} capsids has been shown to have the character of a first-order, activated process \cite{Zlotnick}. This is consistent with the nucleation-and-growth of ``protein-by-protein" assembly models \cite{Zandi2} where capsid proteins diffuse in towards partially assembled protein shells. In the assembly of complete viruses, this process may be enhanced by the presence of the RNA genome molecule \cite{Shklovskii}. However, recent experiments \cite{Garmann} indicate that viral assembly also can have the character of the \textit{cooperative ordering} of a disordered protein-RNA precursor condensate (see Fig. \ref{assembly}). This second assembly mode is the focus of the current paper. 

A theory for the cooperative ordering transition of a shell of capsid
proteins on the surface of a condensed ssRNA globule may be cast in
the language of a theory of an orientational transition by identifying
$\rho(\Omega)$ with the capsid mass density per unit area on a
spherical surface along a direction $\Omega$. The direction is
measured from the center of a sphere of radius $R$, which we identify
as the inner radius of the assembled capsid (see Fig. \ref{assembly}). 

\begin{figure}[htbp]
\begin{center}
\includegraphics[width=3.5in]{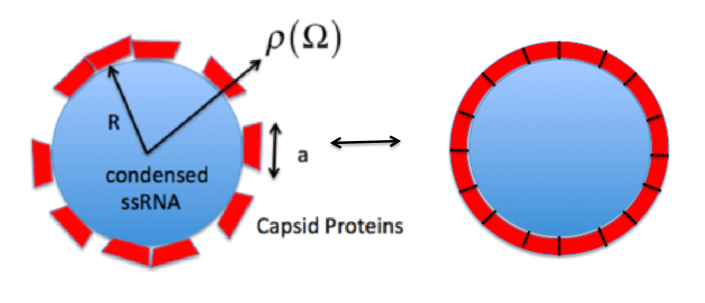}
\caption{Viral assembly by the solidification of a precursor state (left) in the form of a condensate of single-stranded RNA material and groups of capsid proteins. The condensate has a radius $R$ and the structural units have a characteristic size $a$. The mass density of the structural units in a direction $\Omega$ is given by $\rho(\Omega)$.}
\label{assembly}
\end{center}
\end{figure}

The free energy of a Landau theory of orientational transitions is a functional of $\rho(\Omega)$. This ``primary" order-parameter density is the \textit{coarse-grained} density distribution of either the capsid proteins or of groups of capsid proteins that are present in the disordered precursor state as relatively stable entities (see Section V). The full density profile, as measured for example in electron microscopy or X-ray diffraction studies, must be constructed by distributing the proteins, or groups of proteins, in conformity with $\rho(\Omega)$. As discussed in Section V, inclusion in the theory of more of the internal structure of the capsid proteins requires additional ``secondary" order parameters whose values are determined by more complex Landau free energy density functionals with terms that vary from virus to virus, The Landau free energy of the primary order parameter as discussed below can be expected to only represent the more universal aspects.

 In the disordered phase $\rho(\Omega)$ is a constant (equal to $1/(4\pi)$) while in the ordered phase it acquires a density modulations on a length scale $a$ that is of the order of the size of the molecular groups in the precursor state. It follows that if $\rho(\Omega)$ is expanded in spherical harmonics $Y_l^m$ then the characteristic value of $l$ should be of the order of $R/a$. This is in the range of $10-30$ and thus considerably larger than the $l$ values encountered in studies of orientational ordering of liquid crystals \cite{Kats}. 

If one restricts the expansion of $\rho(\Omega)$ to a single value of $l$ in accordance with the tenets of Landau theory, then the icosahedral density modulations can be directly obtained by requiring $\rho(\Omega)$ to be invariant under the actions of the icosahedral group $I$. For given $l$, the resulting linear combination of spherical harmonics $Y_l^m$ are well known as the \textit{icosahedral spherical harmonics}, denoted by $\mathcal{Y}_h(l)$ \cite{Klein}. It is not possible to construct icosahedral spherical harmonics for all $l$ values. An icosahedral spherical harmonic can be constructed for even values of l given by $l$ if $l=6j+10k$ with $j,k\in\{0,1,2,\cdots\}$ and for odd values of $l$ if $l=15+6j+10k$ \cite{golubitsky2012singularities}. For even $l$ less than 14, this restricts $l$ to $6, 10, 12$ while there are no restriction for larger even $l$. Odd $l$ icosahedral states are restricted to $l=15, 21, 25, 27$ for $l$ less than $29$ while there are no restrictions for larger odd $l$. 

The lowest $l$ value that allows construction of an icosahedral shell is thus $l=6$. The properties of the $l=6$ orientational phase transition have already been extensively explored in the context of quasi-crystals and glasses \cite{Steinhardt, Jaric}. A conventional first-order phase transition separates the uniform state from a stable, $l=6$ icosahedral state. The reason  the transition is first-order is because of the presence of a non-zero cubic term in the Landau free energy. The $l=6$ icosahedral state competes with states that have different symmetries, but it is stable over a substantial sector of parameter space. 

\begin{figure}[htbp]
\begin{center}
\includegraphics[width=3.5in]{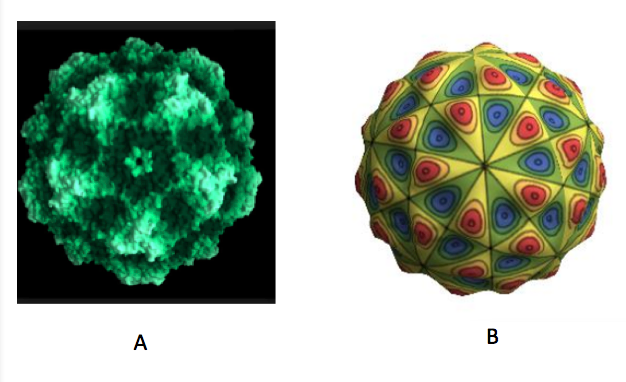}
\caption{A) Reconstruction of the \textit{parvovirus}, viewed along a 5-fold symmetry axis (from ref.\footnote{http://www.virology.net/BigVirology/BVDNAparvo.html}). B) The $\mathcal{Y}_h(15)$ icosahedral spherical harmonic.}
\label{Parvo}
\end{center}
\end{figure}

Could the $l=6$ icosahedral orientational transition be viewed as a model for Landau theories of the assembly of viral capsids? Like all proteins, those comprising a capsid (which are known as ``subunits") are chiral molecules that lack an inversion center. X-ray diffraction and electron microscopy reconstructions of viral capsids sometimes display a pronounced chiral character \cite{baker}. However, $\mathcal{Y}_h(6)$ is even under inversion, as are all even icosahedral spherical harmonics. For this reason, all even $l$ icosahedral spherical harmonics were excluded from the current Landau theory of viral assembly ~\cite{lorman2008landau, lorman2007density}. The absence of inversion symmetry is also the reason that the symmetry group $G_0$ of the disordered phase is $SO(3)$ rather than $O(3)$. The lowest \textit{odd} icosahedral spherical harmonic is $l=15$. This should correspond to the smallest icosahedal viral shells. Fig. \ref{Parvo}(A) shows a reconstruction of the capsid of the parvovirus \cite{Tsao} viewed along a 5-fold symmetry and compares it $\mathcal{Y}_h(15)$ (Fig.\ref{Parvo}B). The parvovirus belongs to the $T=1$ class, which includes the smallest icosahedral viruses composed of $60$ proteins (though larger viruses, such as the picornavirus, also may be classified as $T=1$). 

The $\mathcal{Y}_h(15)$ icosahedral spherical harmonic indeed reproduces the large-scale features of the capsid. Note that both have a chiral character. Reversing the sign of the $\mathcal{Y}_h(15)$ density produces a new density profile that cannot be transformed into the old density profile by a rotation. There are thus two isomeric densities that correspond to $\mathcal{Y}_h(15)$, which is not the case for $\mathcal{Y}_h(6)$. The $\mathcal{Y}_h(15)$ density goes to zero at the 5-fold symmetry sites, which is true for odd $l$ spherical harmonics in general.  A systematic comparison between viral capsids and the odd-$l$
icosahedral spherical harmonics can be found in refs.~\cite{lorman2008landau, lorman2007density}. Larger viruses are associated with $\mathcal{Y}_h(l)$ with larger odd $l$. For example, the intensely-studied CCMV virus, discussed in Section V, is associated with $\mathcal{Y}_h(27)$. In this view, the large-scale features of the density of a viral capsid are determined purely by symmetry. The Landau theory proposed in this paper disagrees with that view.

The \textit{cubic} non-linear terms in the Landau free energy vanishes for a density represented by $\mathcal{Y}_h(l)$ with \textit{l} odd. This, in turn, means that cooperative viral assembly should be a \textit{continuous} transition, at least in mean-field theory \footnote{All results in this paper are restricted to mean-field theory, and may be subject to corrections due to thermal fluctuations}. This is in contrast with the protein-by-protein capsid assembly scenario that encounters equilibrium activation energy barriers that are estimated to be large compared to the thermal energy \cite{Zandi2}. 

One problem with the notion that cooperative assembly might be a continuous transition is that in the limit that the sphere radius goes to infinity, the solidification of chiral molecules on a spherical surface should transform to that of the solidification of an infinite layer of chiral molecules. However, liquids composed of chiral molecules undergo conventional, first-order discontinuous solidification transitions \cite{collins}. Resolving whether capsid assembly is a continuous or a discontinuous transition is one of the motivations of the paper. Another focus of the work reported here concerns the stability of the odd-$l$ icosahedral states. In a recent letter \cite{Sanjay} we showed that in Landau theory the $\mathcal{Y}_h(15)$ state is thermodynamically unstable \footnote{Landau Theory and the Emergence of Chirality in Viral Capsids. arXiv preprint arXiv:1606.02769 (2016).}. Stability can be regained by allowing mixing of the $l=15$ and $l=16$ states, while the ordering transition remains continuous. The order parameter is now a pair of coefficients that provides the weigths of the $\mathcal{Y}_h(15)$ and $\mathcal{Y}_h(16)$ icosahedral spherical harmonics. However, this violates the fundamental tenet of Landau theory that continuous symmetry breaking transitions should be described in terms of a single irrep of $G_0$ \footnote{L. D. Landau and E. M. Lifshitz, “Statistical Physics,” chap. 14, section 136, Pergamon Press, Addison-Wesley Publishing Company, Inc., Reading, Mass., 1958.}.

\section{Achiral Landau-Brazovskii Free Energy.}
We start from the Landau-Brazovskii (LB) free energy for solidification, which has been used to describe ordering transitions in liquid crystals, block co-polymers and other soft-matter systems \cite{brazovskii1975phase, Kats}. The defining feature of an LB free energy is the fact that the static structure factor $S(q)$, as obtained from the quadratic terms in the free energy density, should have a maximum at some $|\vec{q}|=k_0$ that diverges at the mean-field critical point. The version of the LB free energy  $\mathcal{H}_{LB}$ that we use is defined by the free energy density

\begin{equation}
\begin{split}
 \mathcal{H}_{LB}=&\int\left(\frac{1}{2}\Big( (\Delta+k_0^2)\rho\Big)^2+\frac{r}{2}\thinspace\rho^2+\frac{u}{3}\thinspace\rho^3+\frac{v}{4}\thinspace\rho^4\right)\;dS.
  \label{eq:LBEnergy}
\end{split}
\end{equation}
(with $k_BT=1$). The integral is over the surface of a sphere of
radius $R$. The $\Delta$ in the first term is the Laplace-Beltrami
operator defined on the spherical surface while $k_0$ is the
characteristic wavenumber of the density modulation of the ordered
state, comparable to the inverse of the size $a$ of the precursor
molecular groups. The next three terms constitute a Taylor expansion
of the free energy density in powers of $\rho$, with $r$, $u$, and $v$
expansion coefficients, and $v>0$. The coefficient $r$ is the control
parameter of the transition. If one retains only the quadratic terms
in the free energy, then the structure factor $S(q)=<|\rho_q|^2>$ is
proportional to $1/[(k_0^2-q^2)^2+r]$ in the large $R$ limit. As
required, $S(q)$ has a maximum at $q=k_0$ that diverges at $r=0$. For
negative $u$ and positive $v$, the LB free energy describes in the
large $R$ limit a first-order solidification transition of a
two-dimensional fluid into a hexagonal solid, represented as the
superposition of three density waves whose wavevectors are oriented at
$120^o$ with respect to each other and whose magnitude is $k_0$
\cite{Pezzutti}. The LB free energy in Eq. (1) is symmetric under inversion. Section IV discusses the inclusion of chiral terms in the free energy.

The condition for a  density modulation to constitute an extremum of the LB free energy is that the first functional derivative $\frac{\delta{{H}_{LB}(\rho)}}{\delta\rho}$ must vanish, which leads to the Euler-Lagrange equation
\begin{equation}
 \frac{\delta{\mathcal{H}_{LB}(\rho)}}{\delta\rho}=(\Delta+k_0^2)^2\rho + r\rho + u\rho^2+v\rho^3 = 0.
\label{eq:sup_PDE}
\end{equation}
As discussed in Section I, if the uniform phase is invariant under the group SO(3) of rotations in three dimensions then Landau theory instructs us that this density modulation must be associated with a single irrep of SO(3) and hence must be expandable in spherical harmonics belonging to a particular $l$ value. Assume that this $l$ value is known. The collection of $2l+1$ numbers $c_m$ in the expansion \begin{equation}\rho(\Omega)=\sum_{m=-l}^l c_m{{Y}_{l}}^{m}(\Omega)\end{equation} is then the orientational order parameter. The condition that the density is real imposes the additional constraint $c^*_{m}=(-1)^mc_{-m}$. Under these conditions, the LB free energy simplifies to

\begin{equation}
\begin{split}
  \mathcal{H}_{LB}([c_m])/R^2=&\int\left(\frac{t_l}{2}\rho^2+\frac{u}{3}\thinspace\rho^3+\frac{v}{4}\thinspace\rho^4\right)\;d\Omega.
  \label{eq:LBEnergy2}
\end{split}
\end{equation}
Here, $t_l=r+[(k_0R)^2-l(l+1)]^2/R^4$ while the integral is over solid angles $\Omega$. The quantity $t_l$ is the effective reduced temperature for icosahedral ordering in this subspace. The optimal $l$ value is the one that minimizes $t_l$ for given $k_0R$. This leads to $l\simeq k_0R$ so the preferred $l$ value increases linearly with the shell radius $R$. We can divide the $k_0R$ axis into segments $l^2<{k_0R}^2<(l+1)^2$ so that in each segment the associated $t_l$ changes sign with decreasing $r$ before any of the other $t_l$. For example, condensation is dominated by the $l=15$ subspace if $(k_0R)^2$ lies in the interval $15<(k_0R)^2<16$.

Inserting the expansion for $\rho(\Omega)$ in spherical harmonics into
Eq. \ref{eq:LBEnergy2} produces a quartic polynomial in the $2l+1$ expansion coefficients $c_m$. Formal minimization leads to the condition that the $2l+1$ projections of the Euler-Lagrange equation on the spherical harmonic $Y_{lm}$ must be zero:
\begin{equation}
  \langle Y_l^m, \frac{\delta{\mathcal{H}_{LB}(\rho)}}{\delta\rho}\rangle=G_m([c_m])=0,\;-l\leq m\leq l.
  \label{eq:sup_EL}
\end{equation}
where the $G_m([c_m])$ are a set of $2l+$1 cubic polynomials in the $c_m$. The polynomials can be obtained from the \textit{Sattinger algorithm}
\cite{sattinger1978bifurcation}, which uses the ladder operators of quantum mechanics as generators of the Lie algebra of $SO(3)$ in order to generate the most general polynomials in the $2l+1$ variables
($c_{-l},\cdots,c_0,\cdots,c_l$) that are
$SO(3)$-equivariant. One only needs to compute as many integrals of products of spherical harmonics as are required to obtain the coefficients that are left undetermined by the Sattinger algorithm. Alternatively, one can also use the Wigner matrices \cite{Wigner} for integrals over products of spherical harmonics.

Once a solution has been found, thermodynamic stability in the fixed $l$ subspace requires that the eigenvalues of the $(2l+1)$ times $(2l+1)$ stability matrix $\langle Y_l^m|\frac{\delta^2 \mathcal{H}_{LB}}{\delta\rho^2}|Y_l^{m'}\rangle$ (or ``Hessian")  are positive apart from three zero eigenvalues that correspond to global rotations over the three Euler angles.

\subsection{Icosahedral Free Energy Extrema: odd $l$}

First consider the odd $l$ icosahedral spherical harmonics. The smallest odd $l$ value that supports an icosahedral state $l=15$. The invariant icosahedral density is proportional to $\mathcal{Y}_h(15)$ \begin{equation}\begin{split}&\mathcal{Y}_h(15)=\frac{3003}{625} Y_{15}^{-15}(\theta ,\phi )-\frac{33}{625} \sqrt{15834} \mathcal{Y}_{15}^{-10}(\theta
   ,\phi )\\ &-\frac{3}{125} \sqrt{\frac{667667}{5}} Y_{15}^{-5}(\theta ,\phi )-\frac{3}{125}
   \sqrt{\frac{667667}{5}} Y_{15}^5(\theta ,\phi )\\ &+\frac{33}{625} \sqrt{15834}
   Y_{15}^{10}(\theta ,\phi )+\frac{3003}{625} Y_{15}^{15}(\theta ,\phi )\label{Y15}
\end{split}
\end{equation}
Only 6 of the 31 coefficients are non-zero. Note that the condition
$c^*_{m}=(-1)^mc_{-m}$ is satisfied. Note also that $m$ values are
multiples of 5, which makes sense given that an icoshedral density
profile density must have five-fold symmetry axes. The three-fold
symmetry axes of the icosahedron are not evident in this expression
because of the choice of the $z$-axis for the $m$ indices, which lies
along a five-fold axis. Acting on the above expression with a rotation
operator that places the $z$-axis along one of the three-fold symmetry
axes, would highlight three-fold symmetry. There are 31 coupled
equations $\langle{Y_{15}^m},\frac{\partial
  \mathcal{H}_{LB}}{\partial\rho}\rangle=0$ for the $c_m$. The
equations are solved for the $c_m$ corresponding the coefficients of
$\mathcal{Y}_h(15)$ in Eq. \ref{Y15}. The density associated with the $l=15$ icosahedral spherical harmonic is thus an extremum of the LB free energy restricted to the $l=15$ sector. 

Denote the undetermined overall multipicative factor of $\mathcal{Y}_h(15)$ by $\zeta$. Inserting this expression in the LB free energy and setting the derivative with respect to $\zeta$ to zero leads to the equation:
\begin{equation}
  \zeta\Big( t_{15} + 0.789 v \zeta^2\Big)=0,
  \label{eq:sup_char_eq}
\end{equation}
where we set $(k_0R)^2=15\times16$ in the middle of the $l=15$ stability segment. This equation now has the standard form of a second-order Landau phase transition. For $t_{15}>0$, the solution is $\zeta=0$ while for $t_{15}<0$ there are two degenerate solutions:
$
  \zeta = \pm 1.126\frac{\sqrt{-t_{15}}}{\sqrt{v}}.
  \label{eq:sup_l15_sol}
$
The solution pair is related by inversion. The onset of an $l=15$ icosahedral density modulation thus involves spontaneous chiral symmetry breaking. The two solutions will be denoted by D (from dextro) for the plus sign and L (from laevo) for the minus sign.

However, as reported earlier \cite{Sanjay}, when the eigenvalues of the $31\times31$ stability matrix for the $l=15$ state are computed numerically, one finds that the icosahedral state has one three-fold degenerate negative eigenvalue and one four-fold degenerate negative eigenvalue. Similar instabilities are encountered also for the subsequent odd $l$ icosahedral states $l=21$, $l=25$, and $l=27$. For example, the icosahedral spherical harmonic $\mathcal{Y}_h(25)$ has a stability matrix with 51 eigenvalues, 27 of which are negative !

The appearance of negative eigenvalues for $l=15$ can also be demonstrated directly. Perturb about the icosahedral state $\rho=\zeta \mathcal{Y}_{h}(15)+\hat{\rho}$ with the perturbation $\hat\rho$ restricted to the
space of $l=15$ spherical harmonics. The change $\delta \mathcal{H}$ introduced by the perturbation is 
\begin{equation}
\delta \mathcal{H}/R^2 =\int\left(\frac{t_{15}}{2}\hat\rho^2
+|t_{15}|\frac{3}{2}(1.126)^2\mathcal{Y}_{15}^2\hat\rho^2\right)\;d\Omega.
\label{eq:sup_second_var}
\end{equation}
up to quadratic order in $\hat\rho$. Try $\hat{\rho}=Y_{15}^0$. By direct integration of (\ref{eq:sup_second_var}) one obtains $\delta \mathcal{H} = -0.003261 |t_{15}| $, which holds for any non-zero value of $w$. Since  $\delta H$ is negative, the icosahedral state is indeed unstable. 

\begin{figure}
 \includegraphics[width=2.5in]{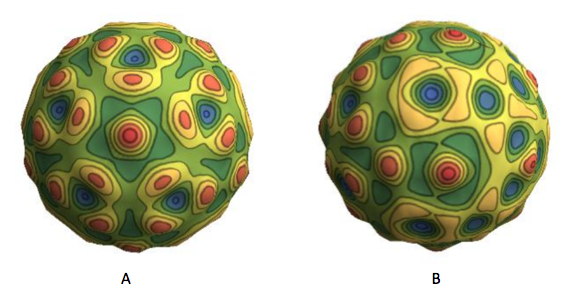}
 \caption{Minimum free energy state in the $l=15$ sector for $t_{15}=-0.1$ and $v=10$. A: View along the single 5-fold symmetry axis. B: View perpendicular to the 5-fold axis. It can be seen from this view that the density is odd under inversion.}
  \label{15}
\end{figure}

When the LB free energy is freely minimized with respect to the $31$
coefficients $c_m$ of the density expanded in $l=15$ spherical
harmonics without imposing icosahedral symmetry, one encounters a
non-icosahedral structure with just a single 5-fold symmetry axis, an
example of which is shown in Fig. \ref{15}.   

\subsection{Icosahedral Free Energy Extrema: even $l$}
\begin{figure}
  \includegraphics[width=2in]{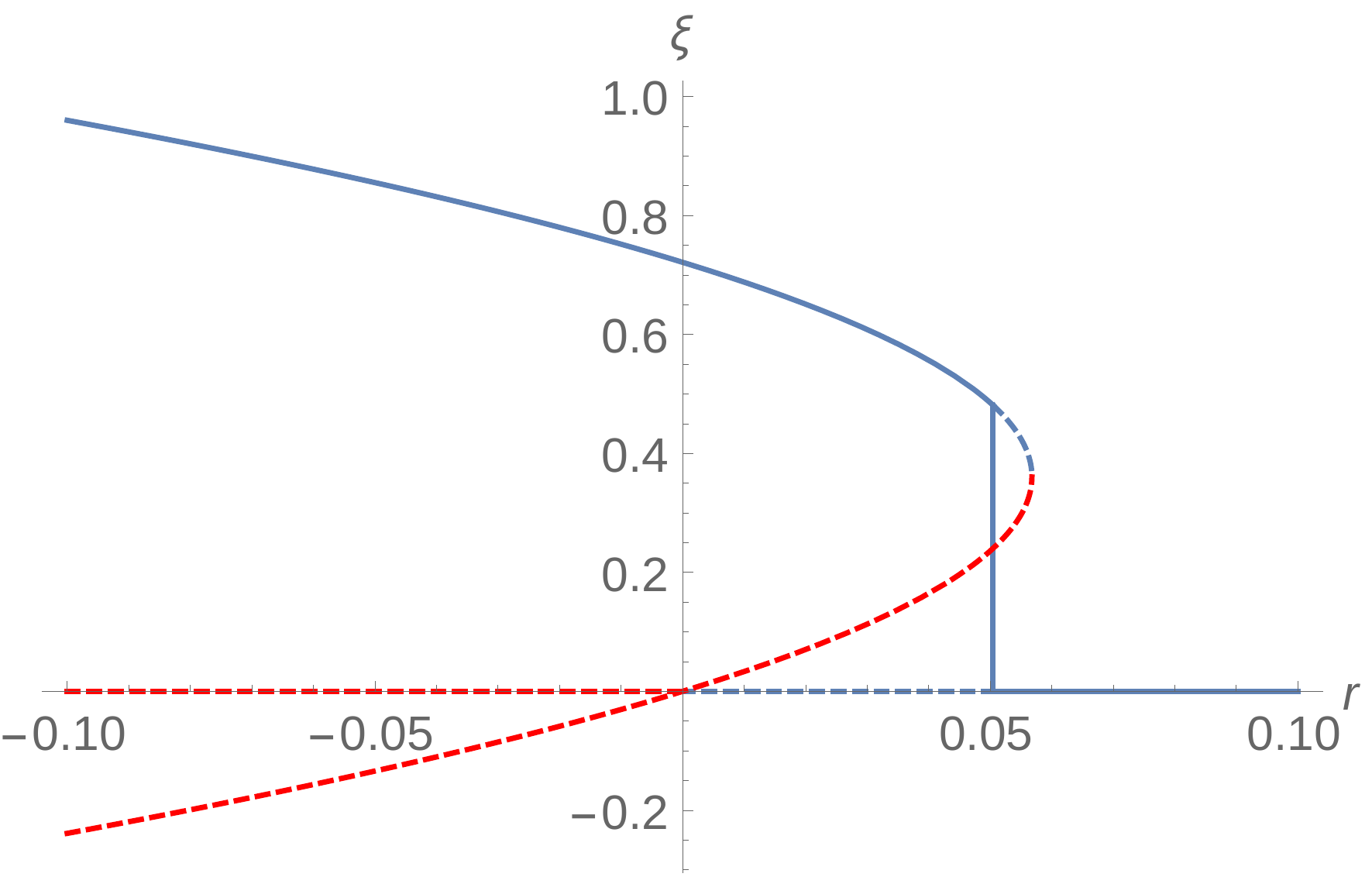}
  \caption{Solution of Eq. \ref{FO} with the order-parameter amplitude $\xi$ as a function of $r$ the instability control parameter. The $k_0$ parameter is set at the optimal value $k_0^2=6\times7$ for the $\mathcal{Y}_h(6)$ icosahedral spherical harmonic. Stable solution branches are shown as a solid blue line, metastable branches as a dashed blue line. Solution branches with one or more negative eigenvalues are shown as a dashed red line.}
  \label{fig:l=6 solution curves}
\end{figure}

It is also useful to examine even $l$ icosahedral spherical harmonics. The lowest even $l$ value is the well-studied $l=6$, with associated icosahedral spherical harmonic $\mathcal{Y}_h(6)=Y_{0,0}+\sqrt{\frac{7}{11}}Y_{6,5}-\sqrt{\frac{7}{11}}Y_{6,-5}$ \footnote{see Appendix A}. Only three of the 13 coefficients $c_m$ are non-zero. The 13 coupled equations $G_m([c_m])=0$ for the expansion coefficients are solved by these $c_m$, which confirms that the icosahedral state is again an extremum of the LB free energy. If we denote the undetermined overall multiplicative factor of $\mathcal{Y}_h(6)$ by $\xi$, insert the Ansatz $\rho_{I}= \xi\mathcal{Y}_h(6)$ into the free energy and minimizing the free energy with respect to $\xi$, one obtains:
\begin{equation}
  r\xi+\tilde{u}\xi^2+\tilde{v}\xi^3=0,
  \label{FO}
\end{equation}
with $\tilde{u}=u(50\sqrt{13}/323\sqrt{\pi})$ and $\tilde{v}=2145v/(1564\pi)$ for $(k_0R)^2=6\times7$, so in the middle of the $l=6$ stability segment. This equation has the standard form for a first-order Landau phase transition. Fig. \ref{fig:l=6 solution curves} shows the order-parameter amplitude $\xi$ as a function of $r$.
Solution branches for which the $13\times13$ stability matrix has positive eigenvalues plus three zero eigenvalues are shown as solid and dashed blue lines. The solid blue line is the minimum free energy state. The onset of icosahedral order as a function of $r$ proceeds via a first-order transition with an order parameter discontinuity. The dashed red curve represents a solution for which the Hessian has negative eigenvalues, indicating that this is a first order transition with metastability. The stability diagram is consistent with the earlier work on $l=6$.

\begin{figure}
 \includegraphics[width=3.5in]{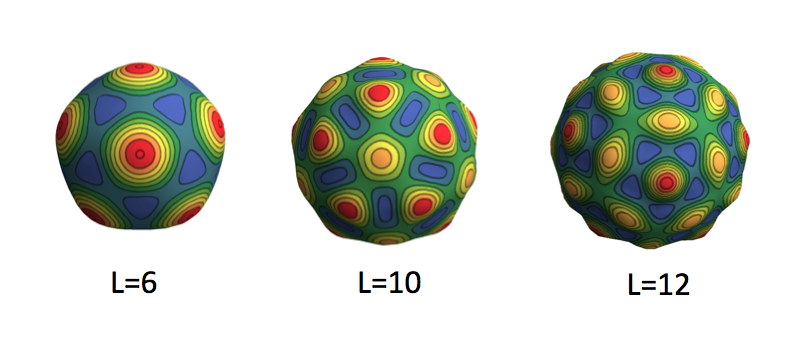}
 \caption{Stable icosahedral density profiles for $l=6$, 10 and 12. In all three cases, the five-fold symmetry sites are density maxima for our choice of the sign of $u$, which holds for all even $l$ icosahedral spherical harmonics. }
  \label{OP}
\end{figure}

\begin{figure}
 \includegraphics[width=2.5in]{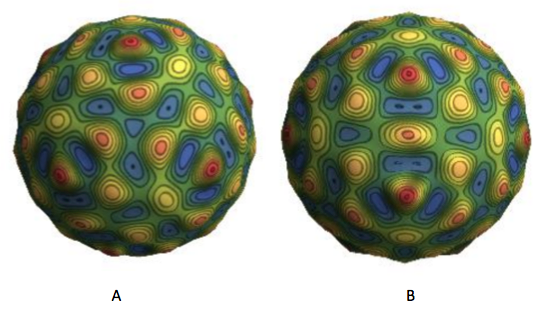}
 \caption{Minimum free energy state in the $l=16$ sector for $t_{16}=-0.1$ and $v=10$ A: View along a 3-fold symmetry axis. B: View along a distorted 4-fold axis.}
  \label{16}
\end{figure}

Stable icosahedral states were found also for $l=10$, $l=12$, and
$l=18$. The first three stable icosahedral density modulations are
shown in Fig. \ref{OP}. 
The Hessian of the $l=16$ icosahedral state has negative eigenvalues
however. The minimum energy structure in the $l=16$ sector is shown in
Fig. \ref{16}.
The structure has only four three-fold axes (see Fig. \ref{16}A) and
no five-fold axes. Surprisingly, it has six distorted
\textit{four-fold} symmetry axes (see Fig. \ref{16}B).  

\subsection{Group theory.}
In order to classify the instabilities of the $l=15$ and $l=16$ icosahedral states, we will reorganize the expansion basis. So far we used the $2l+1$ spherical harmonics $Y_{l}^{m}$ as the basis, which indeed is the natural basis for examining symmetry breaking of the uniform state with $SO(3)$ symmetry. It is not however a convenient basis for examining the icosahedral symmetry breaking encountered in the previous subsection. For that purpose, we construct a new basis composed of groups of linear combinations of $Y_{l}^{m}$ that transform according to the different irreps of $I$ under the symmetry operations of $I$. 

The icosahedral group has five irreps with the character table shown below  
\begin{table}[htp]
\caption{Character table for the five irreducible representations of the icosahedral group}
\begin{center}
\begin{tabular}{|l| c|c|c|c|c|}
 \hline
&   $\, E$ & $\, \mathcal{C}_5, \mathcal{C}_5^4$  & $\, \mathcal{C}_5^2, \mathcal{C}_5^3$ &  $\, \mathcal{C}_2$ &  $\, \mathcal{C}_3, \mathcal{C}_3^2$ \\ 
    \hline
    \hline
   $A_g$  & 1 & 1 & 1 & 1 & 1 \\ 
    $F_{1g}$ & 3 & $\frac{1}{2}(1+\sqrt{5})$ & $\frac{1}{2}(1-\sqrt{5})$ & -1 & 0 \\ 
  $F_{2g}$ & 3 & $\frac{1}{2}(1-\sqrt{5})$  & $\frac{1}{2}(1+\sqrt{5})$ & -1 & 0 \\ 
   $G_g$ & 4 & -1 & -1 & 0 & 1 \\ 
    $H_g$ & 5 & 0 & 0 & 1 & -1 \\
    \hline
\end{tabular}
\end{center}
\label{table:charactertable}
\end{table}%
The notation for the rotational symmetry operations $\mathcal{C}_j$ of $I$ is detailed in Appendix A.  
The second column, which contains the characters associated with the identity $E$, also gives the dimension $d$ of the irrep. Using this character table, one can project any spherical harmonic $Y_{l}^{m}$ onto an irrep $i$ of $I$:
\begin{equation}
\sum_{j=1}^{5} \chi_i^{(j)} \sum_{{ \bf R} \in C_j} Y_l^m({ \bf R}(\widehat{r})) \label{eq:icosa1}
\end{equation}
Here, {\bf R} is one of the 60 symmetry operations of $I$ while $j$ runs over the five entries of the appropriate row of the character table. A rotated spherical harmonic $Y_l^m({ \bf R}(\widehat{r}))$ can be expanded in terms of the $2l+1$ unrotated spherical harmonics with the same $l$:
\begin{equation}
Y_l^m({ \bf R}(\widehat{r}))=\sum_{m'=-l}^{m'=+l}[D_{mm'}^l]^*Y_l^{m'}(\widehat{r})
\end{equation}
where $[D_{mm'}^l]^*$ is the complex conjugate of an element of the (tabulated) Wigner D-matrix \cite{Wigner}. By applying this projection method to any of the $2l+1$ spherical harmonics, one obtains five linear combinations of the spherical harmonics corresponding to the five irreps of $I$, generating in total a set of five times $2l+1$ combinations. This set is over-complete and in some cases the operation in Eq. (8) produces a vanishing result. By diligently sifting through this set and extracting all linearly independent terms, one can construct of a new orthonormal basis composed of $2l+1$ real basis states $\phi_i^{(j)}(\theta, \phi)$. The subscript $i$ again refers to the irrep, while the superscript $j$ in parentheses runs over the d-dimensional basis \textit{within} for that representation. For example, $j$ runs from 1 to 3 for irrep $F_{1g}$.  The normalization is
\begin{equation}
\int_0^{\pi} \int_0^{2 \pi} \phi_i^{(j)}( \theta, \phi)^2  \sin (\theta)d \phi \, d \theta = 1 \label{eq:normeq}
\end{equation}
The basis function corresponding to the one dimensional irrep $A_g$ remains the familiar $\mathcal{Y}_h(l)$, which can be explicitly constructed by this route. In Appendix A we illustrate the method for the familiar case of $l=6$.

\subsection{$l=15$}

Applying this method to the $l=15$ case, one obtains one instance of the one-dimensional irrep ($A_g$ or $\mathcal{Y}_h(15$) two instances of each of the three dimensional representations ($F_{1g}$ and $F_{2g}$), two instances of the four dimensional representation $G_{g}$ and two instances of the five dimensional representation $H_{g}$. This yields a total number of new basis states equal to $1 \times 1 + 2 \times 3 + 2 \times 3 + 2 \times 4 + 2 \times 5 =31$ that equals to $2l+1$ as required for a complete basis. 

Start by allowing only the one-dimensional irrep $A_g$. Including non-linear terms, the LB free energy takes the form we saw before:
\begin{equation}
\frac{t_{15}}{2} \zeta^2 + 0.21\frac{v}{4} \zeta^4  \label{eq:l15freen}
\end{equation}
As discussed, this predicts a continuous transition to an ordered $l=15$ icosahedral state at the point $t_{15}=0$. Now allow the other $30$ basis functions to participate in the Hamiltonian but include their expansion coefficients only to quadratic order. The resulting Hamiltonian is then diagonal in terms of these expansion coefficients plus $\zeta$. Diagonal entries are the same for the basis states of a given irrep. We  define $t_{i}(k)$ to be the quadratic coeffcient for the instance $k$ that irrep $i$ is realized. 

Table \ref{table:icosa3} lists the $t_{i}$s: 
\begin{table}[htp]
\caption{Quadratic coefficients $t_{i}$ for the nine instances of irreps of the icosahedral group for $l=15$. The relationships hold when $t_{15}<0$}
\begin{center}
\begin{tabular}{|l|c|}
\hline
Irreducible representation & Quadratic coefficient, $t_{15,i}$ \\
\hline
\hline
$t_{A_g}(k=1)$ & $2|t|_{15}$ \\
\hline
$t_{F_{1g}}(k=1)$ & 0 \\
\hline
$t_{F_{1g}}(k=2)$ & $0.016385 |t_{15}|$ \\
\hline
$t_{F_{2g}}(k=1)$ & $-0.0184816 |t_{15}|$ \\
\hline
$t_{F_{2g}}(k=2)$ & $0.00332569|t_{15}|$ \\
\hline
$t_{G_{g}}(k=1)$ & $-0.00239087|t_{15}|$ \\
\hline
$t_{G_{g}}(k=2)$& $0.0429669 |t_{15}|$ \\
\hline
$t_{H_{g}}(k=1)$ & $ 0.335413|t_{15}| $ \\
\hline
$t_{H_{g}}(k=2)$ & $ 0.0618115|t_{15}|$ \\
\hline
\end{tabular}
\end{center}
\label{table:icosa3}
\end{table}
All coefficients are proportional to the effective temperature $t_{15}$ of the one-dimensional irrep $A_g$ shown in the first row. The zero in the second row for one of the $F_{1g}$ corresponds to the three zero eigenvalues associated with rotation, as mentioned earlier. Instabilities are associated with negative values for $t_{i}(k)$. The first one is a three-fold instability associated with $F_{2g}$ while the second one is a a four-fold instability associated with $G_g$ so together there are seven negative eigenvalues.

The new basis is complete in the $l=15$ space but for it to be an
\textit{economical} basis for the present case, the structure shown in
Fig. \ref{15} should be the superposition of the one-dimensional
representation $A_g$ plus a small number of densities that transform
according to the higher dimensional irreps. In the simplest case,
those would be the two irreps with negative $t_{i}$. We find that the
structure of Fig. \ref{15} is a linear superposition of $A_g$ plus one
copy each of the four d=3 irreps ${F_{1g}}(k=1,2)$ and
${F_{2g}}(k=1,2)$. In each case, one must pick an eigenvector with
5-fold symmetry.  All four of the five-fold axes must be aligned with
one of the 5-fold axes of $A_g$. The d=4 irrep $G_g$ does not appear
to contribute. Group theory thus indicates that Fig. \ref{15} can be viewed a distorted $l=15$ icosahedral structure. The expectation that only irreps with negative $t_{i}$ should be present in the final structure is, however, wrong.

\subsection{$l=16$}

When the same analysis is carried out for $l = 16$, the lowest even
$l$ icosahedral structure that is unstable, one encounters one
instance of the one dimensional representation, $A_g$, corresponding
to $\mathcal{Y}_h(16)$, two instances of the three dimensional
representation $F_{1g}$, one instance of the three dimensional
representation $F_{2g}$, two instances of the four dimensional
representation $G_g$ and three instances of the five dimensional
representation $H_g$. The total number of basis states is $1 \times 1
+ 2 \times 3 + 1 \times 3 + 2 \times 4 + 3 \times 5 = 33$, which
equals $2l+1$ as required. Restriction to the one dimensional
representation leads to the free energy expression Eq. (\ref{FO}). If the cubic coefficient is equal to zero then there is a continuous ordering transition as $t_{16}$ passes through zero, and the ordered state is quadratically stable. However, when the third order coefficient is non-zero the ordered state acquires instabilities. Fig. \ref{fig:l16stabplot} is a plot of quadratic coefficients:
\begin{figure}[htbp]
\begin{center}
\includegraphics[width=2in]{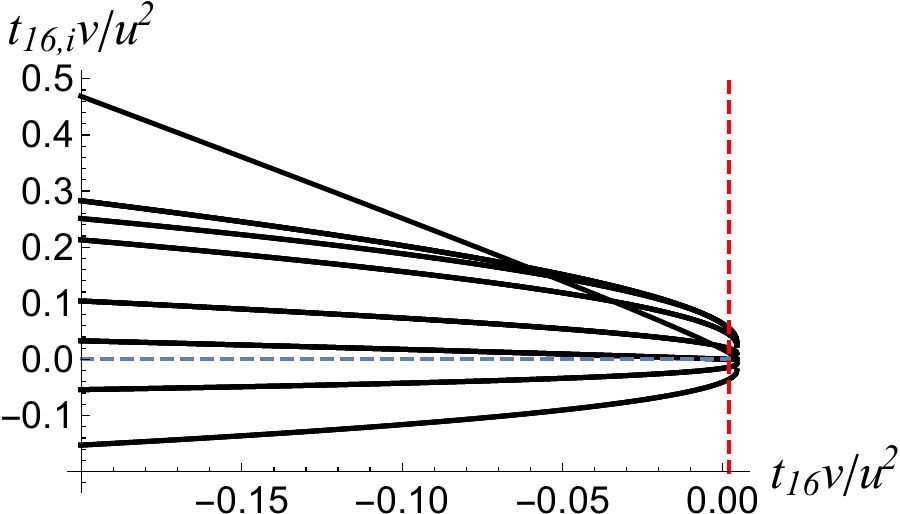}
\caption{The quadratic coefficients, $t_{16,i}$ when there is ordering for $l=16$. The nine curves correspond to the nine irreducible representations comprising the basis set (one one dimensional, three three dimensional, two four dimensional and three five dimensional). The dashed horizontal curve along the axis $t_{16,i}v/u^2=0$, corresponds to one of the three dimensional representations that reflects the rotational invariance of the energy of the condensed state. The vertical dashed line indicates the onset of the first order transition to ordering at $t_{16}\simeq0.002u^2/v$. All curves terminate at the far right, corresponding to the onset of a local non-zero minimum in the free energy at $t_{16}\simeq 0.004 u^2/v$.  }
\label{fig:l16stabplot}
\end{center}
\end{figure}
Two of the curves lie below the horizontal axis and correspond to instabilities, one of them to a three dimensional irrep and the other to a four dimensional irrep. One of the curves lies along the horizontal axis, corresponding to a three dimensional irrep and reflecting the rotational invariance of the energy of the condensed state. 

This basis set applied to the decomposition of the $l=16$ structure
shown in Fig. \ref{16} does \textit{not} provide an economic
description. Instead, Fig. \ref{16} comports with the symmetry of the \textit{octahedral group}. The octahedral group is composed of the 24 rotations that leave the octahedron---or its dual regular polyhedron, the cube---unchanged. This group has five irreducible representations, the character table of which is shown in Table \ref{table:octahedral}. 
\begin{table}[htp]
\caption{The character table of the octahedral group. As in the case of Table \ref{table:charactertable}, the number in the second second column is the dimensionality of the representation.}
\begin{center}
\begin{tabular}{l|ccccc}
    \hline
    Irrep & $E$ &$\mathcal{C}_3(8)$ & $\mathcal{C}_2(3)$ & $\mathcal{C}_2(6)$ & $S_4(6)$ \\ 
    \hline
    \hline
    $A_1$ & 1 & 1 & 1 & 1 & 1 \\ 
    $A_2$ & 1 & 1 & 1 & -1 & -1 \\ 
    $E$ & 2 & -1& 2 & 0 & 0 \\ 
   $F_2$ & 3 & 0 & -1 & 1 & -1 \\ 
   $F_1$ & �3& 0 & -1 & -1 & 1 \\ 
    \hline
\end{tabular}
\end{center}
\label{table:octahedral}
\end{table}
In the case of this group there are two one dimensional irreps. The first, $A_!$, is invariant under the action of rotations that leave the octahedron unchanged. The second, $A_2$, is either transformed into itself or minus itself under such a rotation. 

For the case of $l=16$ there are two instances of the irrep $A_!$ and
one instance of $A_2$, as well as three of the two dimensional irrep
$E$ and four each of $F_1$ and $F_2$, for a total of $2 \times 1 + 1
\times 1 + 3 \times 2 + 4 \times 3 + 4 \times 3$ = 33 basis states,
sufficient to replace the 33 $Y_{16}^m(\theta, \phi)$'s as a complete
basis set.  We find that the structure of Fig. \ref{16}, which was obtained by minimization of the Landau-Brazovskii free energy, consists of a linear combination of  three densities corresponding to two instances of $A_1$ and one instance of $A_2$. The structure thus can be described as having distorted octahedral or cubic symmetry.

\section{Landau Theory beyond Landau: icosahedral order in the $l=15+16$ composition space}

We found that the minimum free energy state in the $l=15$ sector is a distorted icosahedral structure with just a single five-fold symmetry axis. Current Landau theory thus cannot account for the stability of, say, the icosahedral parvovirus whose capsid has twelve five-fold axes with a density that resembles the $l=15$ icosahedral spherical harmonic $\mathcal{Y}_{15}(h)$. In this section we go beyond current Landau theory by allowing the order-parameter to be described \textit{in an essential way} by more than one irrep. In current Landau theory, it already is possible for a primary order parameter associated with one irrep to entrain a secondary order-parameter associated with a different irrep. This is possible when the non-linear terms in the Landau energy produce terms that are linear in the secondary order parameter times an (integer) power of the primary order parameter. Alternatively, at an accidental degeneracy point the uniform state may lose stability against two different irreps when both reduced temperature and another parameter are being varied. This leads to \textit{multicritical} behavior at that specific point of the phase diagram with different critical exponents. With ``essential" we exclude these two cases but refer to the fact that an ordered state with a particular symmetry \textit{only exists} as the superposition of multiple irreps. 

Mixing just two states with different $l$ values is natural near the points along the $k_0 R$ axis that mark the borders between the stability segments of $l$ and $l+1$. For $l=15$ and $l=16$, the point where $t_{15}=t_{16}$ is at $k_0R=16$. More generally, if an odd $l$ segment supports an icosahedral state then $l$ can be expressed as $l=15+6j+10k$ for certain integers $j$ and $k$. The adjacent segment at $l+1=6(j+1)+10(k+1)$ then necessarily obeys the condition for an even $l$ segment to be able to support an icosahedral state. Every odd $l$ segment that supports an icosahedral state is thus bordered at $k_0R = l+1$ by an even $l+1$ segment that also supports an icosahedral state.  Note that this is \textit{not} the case for even $l$: the $l=6$, $l=10$, and $l=12$ icosahedral states do not have icosahedral neighbors.

\subsection{Pairs of icosahedral order parameters.}

Consider the space formed by the composition of the $l=15$ and $l=16$ subspaces. In this extended space, icosahedral density modulations can be expressed as
\begin{equation}
  \rho(\zeta,\xi) = \zeta \mathcal{Y}_h(15) + \xi \mathcal{Y}_h (16),
  \label{eq:coupled_ansatz}
\end{equation}
This density is characterized by the \textit{pair} of order-parameter amplitudes $\zeta$ and $\chi$. When this ansatz is inserted into the LB free energy density and minimized with respect to the pair $(\zeta,\xi)$ then the free energy is found to have an extremum when $(\zeta,\xi)$ obeys the pair of coupled cubic equations \footnote{ $t_{16}=((k_0R)^2-16\times17)^2+r$, $t_{15}=((k_0R)^2-15\times16)^2+r$, $a_1/16=-80.4$, $a_2/16=-3084.1$, $u_1=0.13494 u$, $u_2=0.234946 u$, $v_1=1.04204 w$, $v2=0.681228 w$, $u_3=0.623577 u$, $v_3=0.789107 w$, and $v_4=0.904033 w$}:
\begin{subequations}
  \begin{equation}
    t_{16}\xi + u_1 \xi^2 + u_2\zeta^2 + v_1 \xi^3+ v_2 \xi\zeta^2 = 0,
    \label{xi}
  \end{equation}
  \begin{equation}
    t_{15}\zeta + u_3\xi\zeta + v_3 \zeta^3 + v_4\zeta \xi^2 = 0,
    \label{zeta}
  \end{equation}
  \label{zeta}
\end{subequations}
A numerical solution of the pair of coupled equations at $k_0R=16$ is shown in Fig. \ref{15x16(3)}:
\begin{figure}
 \includegraphics[width=3.0in]{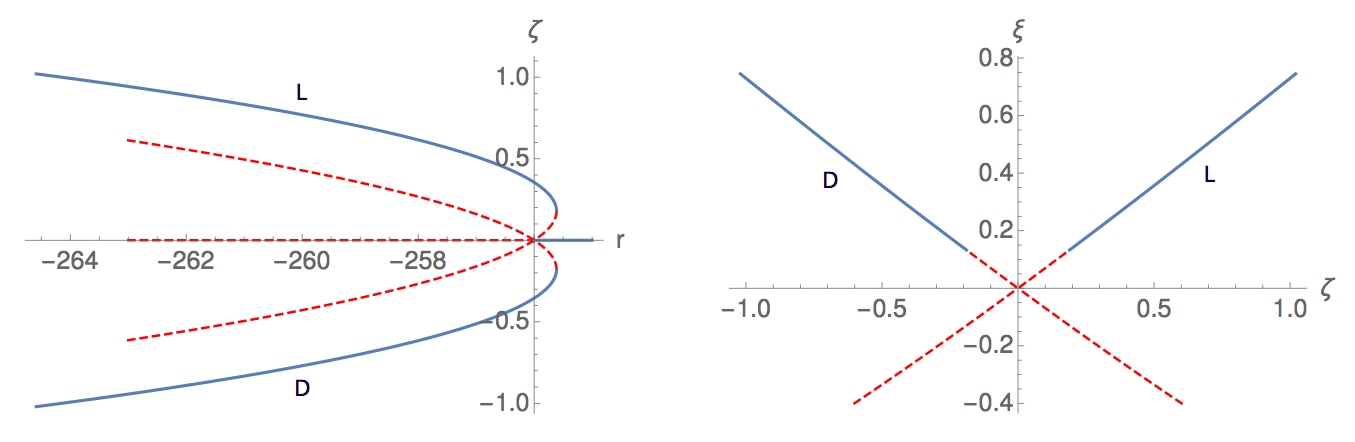}
 \caption{Solution branches of Eq. \ref{zeta} for $k_0=16$, $u$ = -10, $w=10$ that have stable sections. Stable sections of the branch are shown in blue, unstable sections in red.  (left) Coefficient $\zeta$ of $\mathcal{Y}_h(15)$ as a function of the control parameter $r$. The two branches marked $D$ and $L$ are degenerate. (right) Coefficient $\xi$ of $\mathcal{Y}_h(16)$ as a function of $\zeta$ for varying $r$.} 
\label{15x16(3)}
\end{figure}
Fig. \ref{15x16(3)} shows the order-parameter pair $\zeta,\xi$ for two
degenerate solution branches, marked $D$ and $L$ that are related
under inversion when $\zeta\rightarrow-\zeta$ and
$\xi\rightarrow\xi$. The superposition state is thus neither even nor
odd under inversion. Note the resemblance of Fig. \ref{15x16(3)}A with
Fig. 5, suggestive of a first-order transition. Fig. \ref{15x16(3)}B shows that $\zeta$ and $\xi$ are comparable in magnitude and (approximately) proportional to each other.

\subsection{Stability Diagrams.}

\begin{figure}
 \includegraphics[width=3.5in]{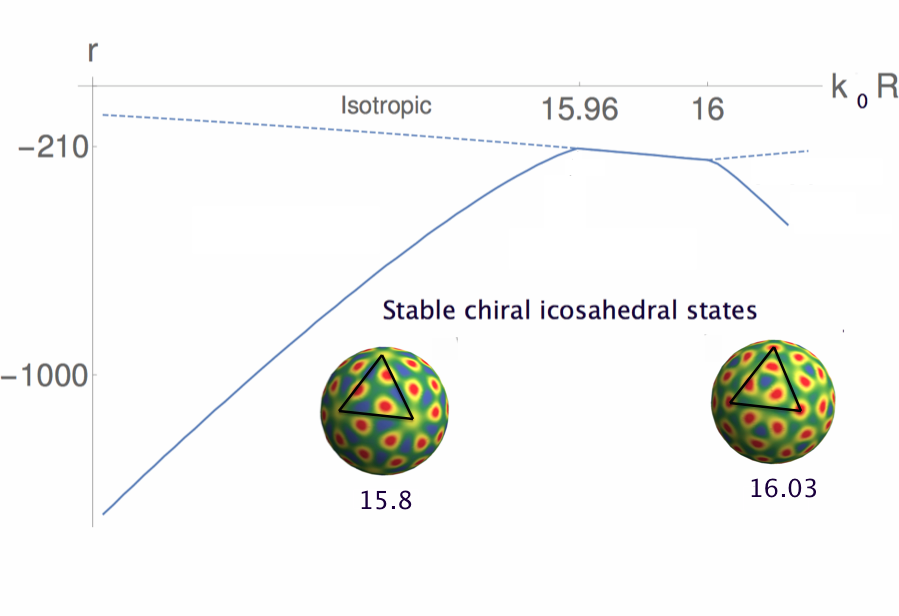}
 \caption{Stability diagram of the $l=15+16$ superposition state in the $r,k_0R$ plane near $k_0R=16$ with $r$ the global control parameter. Solid blue lines: locus of points where the superposition icosahedral state acquires negative eigenvalues. Dashed lines: locus of points where the uniform state acquires negative eigenvalues. The densities of two states with different $k_0R$ at the opposite stability limits are shown. } 
\label{PD} 
\end{figure}

The stability of the $15+16$ superposition state was examined, as
before, by computing the eigenvalues of the $64\times64$ Hessian
matrix. In the $r,k_0R$ plane there is an area near $k_0R=16$ where
the superpositions state has positive eigenvalues plus the three zero
eigenvalues, as shown in Fig. \ref{PD}. 
The icosahedral state has no negative eigenvalues inside the area bordered by the solid blue lines. The diagram shows that if, for fixed $r$, the dimensionless radius of curvature $k_0R$ is increased from the lower stability to the upper stability limit then the icosahedral state changes from ``15-like", with 60 maxima, to ``16-like" with 72 maxima with the new maxima appearing at the twelve 5-fold sites. The stability of icosahedral states around $k_0R=16$ could be conceived as a form of ``interference" between the $l=15$ minimum energy state, which has a single five-fold axis, and the $l=16$ minimum energy state which has four three-fold axes. The icosahedral state is then a compromise that allows both types of symmetry to coexist in one structure. The dashed lines give the locus of points where the uniform state acquires negative eigenvalues. The dashed and solid lines coincide along the central section of the blue lines, which means that along this section there could be a continuous transition from a uniform to an icosahedral state.

\begin{figure}
 \includegraphics[width=3.5in]{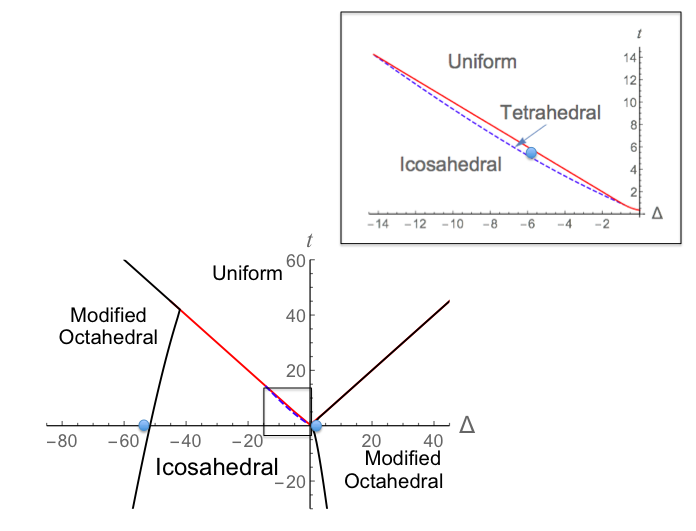}
 \caption{Enlarged view of the central section of the stability diagram of the $l=15+16$ superposition state near $k_0R=16$. The variables $t$ and $\Delta$ are defined by the conditions that $t_{15}=t-\Delta$ and $t_{16}=t+\delta$. The two non-icosahedral modulated states have here modified octahedral symmetry. The two black lines bordering the icosahedral sector are \textit{spinodal lines} with lines of first-order transitions tracking the spinodal lines just inside the icosahedral sector (not shown). The red line \textit{outside} the boxed sector indicates a line of continuous phase transitions between the isotropic and icosahedral states. The inset shows an enlarged view of the boxed sector when a tetrahedral state interposed between the isotropic and icosahedral states. Here, the red line is the stability limit of the uniform state. Blue dots: states whose densities are shown below.} 
\label{PD2} 
\end{figure}

\begin{figure}
\includegraphics[width=1.5in]{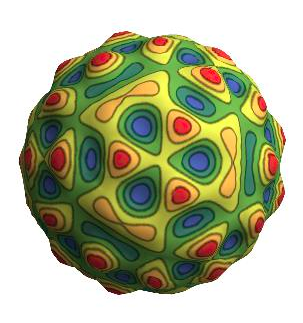}
\caption{Tetrahedral density corresponding to  $\Delta=-6$ and $t=-5.7$, inside the tetrahedral strip in the stability diagram shown along one of the three-fold axes (blue dot)}
\label{tetra} 
\end{figure}

Enlarged views of the central section of the stability diagram are
shown in Fig. \ref{PD2}
We introduced here the new variables $t$ and $\Delta$ such that the
reduced temperature in the $l=15$ sector equals $t_{15}=t-\Delta$ and
that in the $l=16$ sector $t_{16}=t+\Delta$. The linear stability
thresholds $t_{15}=0$ and $t_{16}=0$ of the uniform phase are thus
$t=\Delta$, respectively, $t=-\Delta$. In terms of $r$, $k_0$ and $R$
the new variables can be expressed as
$t=r+[(k_0R)^4-256(k_0R)^2+65792]/R^4$ and
$\Delta=[-16(k_0R)^2+8192]/R^4$. In Fig. \ref{PD2}, the solid red line
is the stability threshold of the uniform phase. The inset shows that
in the narrow sliver, between the solid red and dashed blue lines, the
minimum free energy state has \textit{tetrahedral} symmetry, an
example of which is shown in Fig. \ref{tetra}
The tetrahedral density is shown along one of the four triangular faces. The tetrahedral sliver disappears near $\Delta=-14$.

\subsection{Phase Transitions.}
By combining the stability diagram with numerical calculation of the free energy and the order parameters we can relate the stability diagram to a phase diagram in the $t-\Delta$ plane. 

\begin{figure}
\includegraphics[width=2.3in]{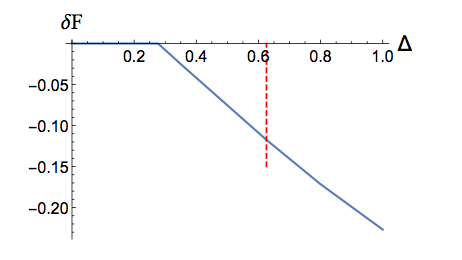}
\caption{Free energy difference $\delta F$ between the lowest free energy state and the icosahedral free energy as a function of $\Delta$ for fixed $t=-0.1$. The red dashed line shows the stability limit of the icosahedral state}
\label{SD1} 
\end{figure}

We start with the solid black line for positive $\Delta$ and negative
$t$. The minimum free energy state in the non-icosahedral sector on
the right side of the icosahedral sector is identical to the
modified-octahedral $l=16$ state that was discussed in the previous
section. The density corresponding to the blue dot in that sector,
near $t=\Delta=0$ in Fig. \ref{PD2} was already shown in Fig. \ref{16}. Fig. \ref{SD1} show the difference $\delta F$ between the lowest free energy state and the icosahedral free energy as a function of $\Delta$ for fixed $t=-0.1$

For $\Delta\lesssim0.3$ the free energy difference is zero, meaning that the icosahedral state has the lowest free energy. For $\Delta\gtrsim0.3$, the modified octahedral state has a lower free energy with 
$\delta F$ going continuously to zero. This is the thermodynamic signature of a first-order phase transition. The icosahedral state acquires negative eigenvalues at the red dashed line, which thus corresponds mathematically to a saddle-point and physically to a \textit{spinodal point}. The solid black line in the stability diagram thus should be interpreted as a spinodal line. A line of first-order transitions runs along the spinodal line. 

We next consider the solid red line outside the boxed sector in
Fig. \ref{PD2} with $\Delta\lesssim-14$. It denotes the joint
stability limits of the icosahedral and uniform states. It is a line
of \textit{continuous phase transitions}, notwithstanding the fact
that the cubic non-linearity in the LB free energy is non-zero! It is
thus indeed possible to have a continuous transition from a uniform
state to an $l=15$-like icosahedral state. We saw that for
$-14\lesssim\Delta<0$, a tetrahedral state interposes between the
icosahedral and uniform states. The group-theoretical method developed
in the previous section to describe instabilities of the icosahedral
state can be readily extended to the $l=15+16$ composition
space. Using this method, we find that the tetrahedral state is a
superposition of the two one-dimensional irreps $A_g$ of the $l=15$
and $l=16$ subspaces plus one copy each of the two four-dimensional
irreps $G_g$ belonging to the $l=15$ and $l=16$ subspaces. For the
case of the tetrahedral state shown in Fig. \ref{tetra}, $A_g$
contributes a fraction of about $0.79$ to the density while $G_g$
contributes the remainder. Transitions from an icosahedral state to a
tetrahedral state of this type have been shown to be first-order
\cite{Hoyle}. The dashed blue inside the boxed sector is thus a line
of first-order phase transitions, as confirmed by Fig. 9. The point where the dashed blue line merges with the red line presumably corresponds to a \textit{tricritical point} but we did not investigate this further.

\begin{figure}
\includegraphics[width=1.5in]{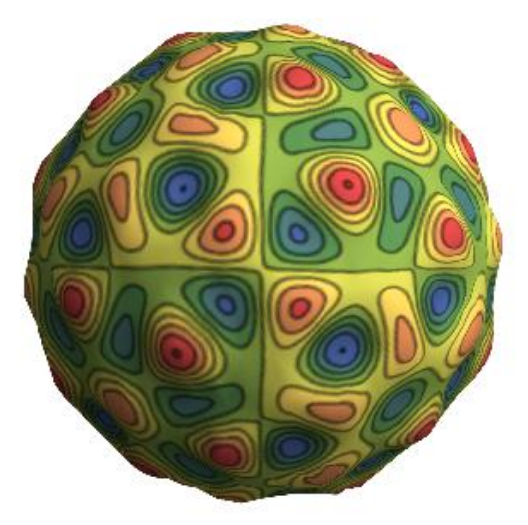}
\caption{Modified octahedral state corresponding to $\Delta = -65$ and
  $t=0.2$ (left blue dot in Fig. \ref{PD2}).}
\label{MO} 
\end{figure}

Finally, the solid black line for negative $\Delta$ and negative $t$
is a mirror image of the one for positive $\Delta$. It is again a
spinodal line where the icosahedral state becomes unstable. A line of
first-order transitions tracks the spinodal line. But there is a
surprise: the minimum free energy state is \textit{not} the
modified-icosahedral $l=15$ state that might have been
expected. Instead, it \textit{also} is a modified-octahedral state as
shown in Fig. \ref{MO}.
This asymmetry is related to the fact that $l=15$ ordering entrains $l=16$ as a secondary order parameter through the cubic term in the free energy, which generates terms that are linear in $l=16$ and bilinear in $l=15$. On the other hand, $l=16$ ordering can not entrain $l=15$ as a secondary order parameter because $l=15$ is odd under inversion. The non-linear terms in the free energy now can not produce terms that are linear in $l=15$. 

The entrainment of even a small amount of $l=16$ has dramatic effects for $\Delta=-65$. If the density is decomposed as
\begin{equation}
\rho( \theta , \phi) = \rho_{15} ( \theta, \phi) + \rho_{16}( \theta \phi) \label{eq:comp11}
\end{equation}
then the ratio $<\rho_{16}^2>/<\rho_{15}^2>$ of the square interated
densitues in Fig. \ref{MO} is only about $2\times10^{-3}$. The density is principally made up of contributions from $l=15$ spherical harmonics, as one should expect. Yet without this admixture, the state would switch back from modified octahedral to modified icosahedral state. The modified icosahedral state is recovered for $\Delta\simeq-400$

Further decomposition in terms of the irreps of the octahedral group using the methods discuss in the previous section yields
\begin{eqnarray}
\rho_{15}( \theta , \phi) &=& \rho_{15, A_1}( \theta , \phi) + \rho_{15, E}( \theta , \phi) + \rho_{15, F_2}( \theta , \phi) \label{eq:comp15} \\
\rho_{16}( \theta , \phi) &=& \rho_{15, A_2}( \theta , \phi) + \rho_{15, E}( \theta , \phi) + \rho_{15, F_1}( \theta , \phi) \label{eq:comp16}
\end{eqnarray}
where the symbols in the subscripts refer to the one dimensional, ($A_1$ and $A_2$), two dimensional (E) and three dimensional ($F_1$ and $F_2$) representations of the octahedral group. The relative square integrated densities contributing to the $\rho_{15}$ term are
\begin{eqnarray}
\frac{ <\rho_{15, A_1}^2>}{<\rho_{15}^2>}= 0.0647066 \label{eq:comp17} \\
\frac{ <\rho_{15, E}^2>}{<\rho_{15}^2>} = 0.326297  \label{eq:comp18} \\
\frac{ <\rho_{15,F_2}^2>}{<\rho_{15}^2>} = 0.608996   \label{eq:comp19}
\end{eqnarray}
Note the small relative weight of the one-dimensional irrep $A_1$. The density deviates strongly from a purely octahedral structure. The relative square integrated densities contributing to the $\rho_{16}$ term are
\begin{eqnarray}
\frac{ <\rho_{16, A_2}^2>}{<\rho_{16}^2>}= 0.114579 \label{eq:comp17} \\
\frac{ <\rho_{16, E}^2>}{<\rho_{16}^2>} = 0.231169   \label{eq:comp18} \\
\frac{ <\rho_{15,F_2}^2>}{<\rho_{15}^2>} = 0.654252   \label{eq:comp19}
\end{eqnarray}
This is completely different from the $l=16$ deformed octahedral state one encounters on the right hand side of the phase diagram, which was a combination of the two one-dimensional irreps $A_!$ and $A_2$.

\subsection{Spinodal lines and group theory}

The instabilities of the $l=15+16$ icosahedral state along the two spinodal lines can be analyzed by group theory. The appearance along this line of three negative eigenvalue is associated with the irrep $F_{2g}$. A natural choice for the eigenvectors is one with the $(1,1,1)$ direction along a three-fold symmetry direction and the $(0,0,1)$ direction along a two-fold direction. Define vectors $\vec{\eta}=(\eta_1,\eta_2,\eta_3)$ with respect to these axes. On general group-theoretic grounds, the Landau energy $\delta F(\vec{\eta})$ in this space must have the form \cite{Hoyle}:
\begin{equation}
\begin{split}
&\delta F(\vec{\eta}) \propto \frac{\lambda_3}{2}|\eta|^2+\frac{V}{4}|\eta|^4+\frac{W}{6}|\eta|^6+\Delta\frac{\sqrt{5}}{2}\eta_1^2\eta_2^2\eta_3^2\\&+\Delta\frac{\sqrt{5}}{60}(\eta_1^6+\eta_2^6+\eta_3^6)+\frac{\Delta}{4}\left(\eta_1^4(\eta_3^2-\eta_2^2)+cyclic\medspace perm.\right)
\label{ISM}
\end{split}
\end{equation}
to sixth order in $\eta$. Here, $\lambda_3$ is the eigenvalue that
changes sign at the transition while $V$ and $W$ are positive
constants. The terms proportional to $W$ and $\Delta$ in Eq. \ref{ISM} are generated by sixth and higher order terms in $\rho$ in the free energy functional that we did not include. $\delta F(\vec{\eta})$ describes a continuous symmetry-breaking transition that takes place at $\lambda_3=0$. For $\Delta=0$, the Landau energy is $O(3)$ isotropic with arbitrary rotations in $\vec{\eta}$ space connecting degenerate states. For $\Delta$ negative, the minimum of $\delta F(\vec{\eta})$ lies along the $C_3$ direction and for $\Delta$ positive along the $C_5$ direction. 

It is suggestive that the two non-icosahedral states bordering the icosahedral state with 5-fold symmetry on the ``left" side and 3-fold symmetry on the ``right" side (see ref. 23) are described by the two minima. In fact, we proposed this to be the case in our first publication \cite{Sanjay}. However, a quantitative check revealed that the two non-icosahedral states bordering the icosahedral state are not well described by the $C_5$ and $C_3$ eigenvectors. When, at the spinodal point, the amplitude of $\vec{\eta}$ begins to grow, it entrains other irreps and the system evolves to a rather different state with modified octahedral symmetry. 

\subsection{$l=25+26$}

\begin{figure}
\includegraphics[width=1.5in]{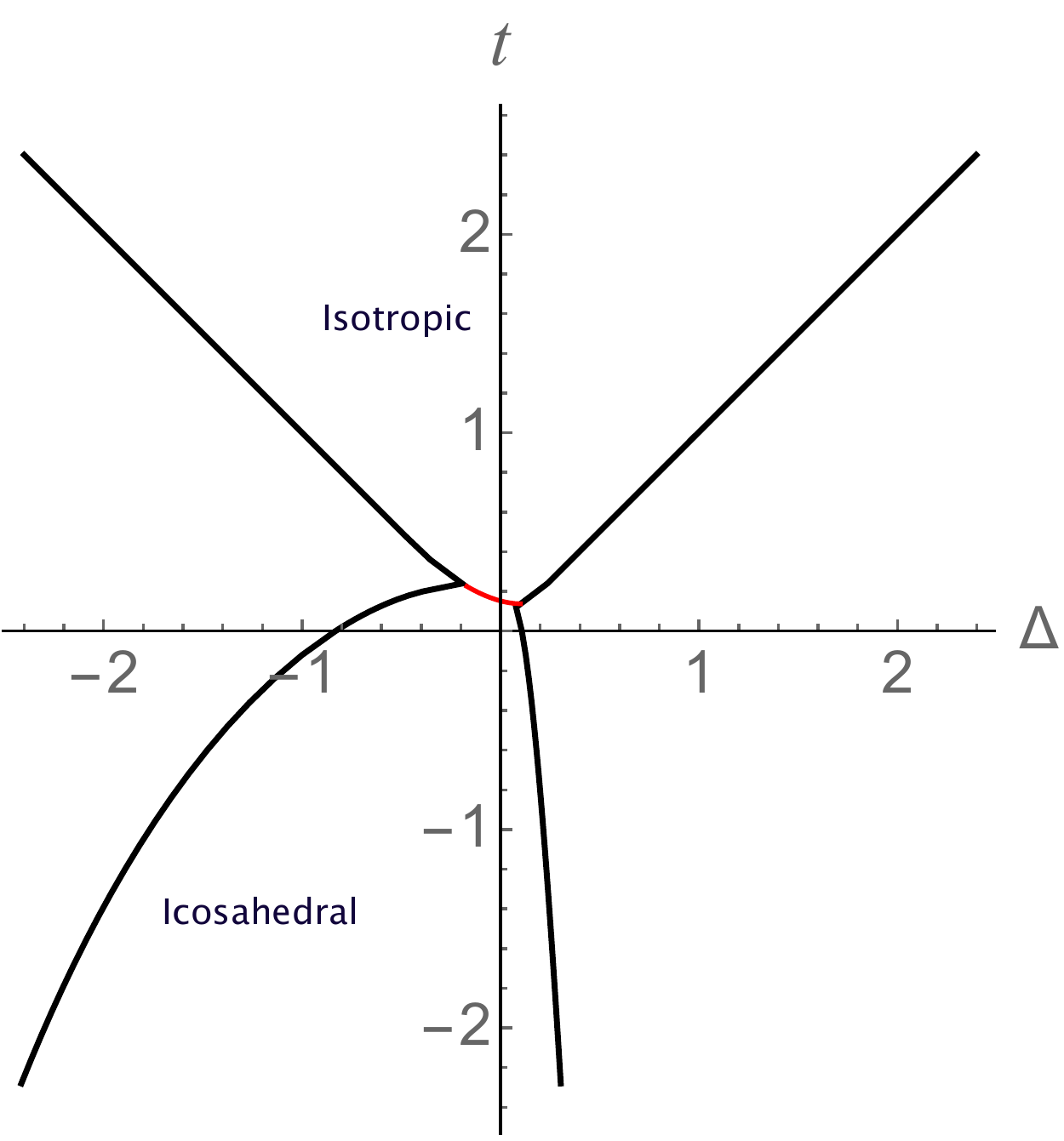}
\caption{Stability diagram in the $l=25+26$ superposition space. The horizontal axis $\Delta$ and the vertical axis $t$ are defined through $t_{25}=t-\Delta$ and $t_{25}=t-\Delta$. The red line is a line of first-order phase transitions from the isotropic to the icosahedral state. The two other lines bordering the icosahedral phase are spinodal lines.}
\label{25+26} 
\end{figure}

We applied the same methods to the $l=25+26$ compositions space. Both
$l=25$ and $l=26$ support icosahedral spherical harmonics but in
neither case is this state stable within the LB free energy. We found
that around $k_0R=26$ a mixed $l=25+26$ icosahedral state can again be
stable. The stability diagram, shown in Fig. \ref{25+26}
is quite similar to the stability diagram in the $l=15+16$. However, the red line separating the icosahedral state from the uniform state now is a line of first-order transitions.

\section{Chiral Landau-Brazovskii Free Energy.}
In the $l=15+16$ composition space, chiral symmetry is broken at the transition between the uniform state and either the $D$ or the $L$ state  However, no such chiral symmetry breaking can take place during viral assembly because capsid proteins have intrinsic chirality. Whether the $D$ or the $L$ icosahedral state is realized is determined by the chirality of the uniform state. 
In this section we will construct a chiral version of the LB free energy that can be applied to ordering transitions in systems composed of chiral units.  Landau free energies for chiral materials, such as cholesteric liquid crystals \cite{deGennes}, are based on obtaining the lowest-order energy density depending on $\rho$ that transforms under inversion as a pseudoscalar density. This pseudoscalar density, multiplied by a pseudoscalar coefficient, is then added to the achiral free energy density constructed from scalar densities.

We start by constructing the pseudoscalars in the large $R$ limit -- so for
a flat plane -- with $\rho$ a scalar density expressed in terms of the cartesian
coordinates $\mathbf{x}=(x,y)$. The chiral contribution to the free energy density $f$ is, as usual,
assumed to be local and to depend on $\rho(\mathbf{x})$ and its first
and second derivatives $\nabla\rho$ and $\nabla\nabla\rho$, respectively. The chiral contribution to the free energy is
\begin{equation}
\Delta\mathcal{H}_{\chi}=\int f(\rho,\nabla\rho,\nabla\nabla\rho)\;d^2\mathbf{x},
\label{eq:sup_free energy}
\end{equation}
The symmetry group of the uniform phase in the large $R$ limit is $SO(2)=\{\mathbf{R}_\theta|\theta\in [0,2\pi) \}$, the
group of proper rotations of the plane. The action on
$\rho$ is defined by
$$\rho(\mathbf{x})\mapsto \rho(\mathbf{R}_\theta^T\mathbf{x})=\rho(\tilde{\mathbf{x}}),$$
where $\tilde{\mathbf{x}}=\mathbf{R}_\theta^T\mathbf{x}$ and
$$\mathbf{R}_\theta = \left(\begin{array}{cc}\cos\theta & -\sin\theta
  \\ \sin\theta & \cos\theta \end{array} \right).$$
The gradient vector $\nabla\rho$ and the tensor $\nabla\nabla\rho$ of
second derivatives transform covariantly:
$$\nabla\rho \mapsto \mathbf{R}_\theta \tilde{\nabla}{\rho},$$
$$\nabla\nabla\rho \mapsto \mathbf{R}_\theta (\tilde\nabla\tilde\nabla
{\rho})\mathbf{R}^T_\theta,$$
where $\tilde{\nabla}\rho = (\partial_{\tilde{x}}\rho,\partial_{\tilde{y}}\rho)$.

The free energy (\ref{eq:sup_free energy}) is invariant under $SO(2)$
if the free energy density satisfies,
\begin{equation}
  f(\rho,\nabla\rho,\nabla\nabla\rho)=f(\rho,\mathbf{R}_\theta\nabla\rho,\mathbf{R}_\theta\nabla\nabla\rho\mathbf{R}_\theta^T), 
\label{eq:sup_inv_cond}
\end{equation}
for all $\theta\in [0,2\pi]$. Expand $f$ as a polynomial in the six 
variables $\rho$, $\rho_x$, $\rho_y$, $\rho_{xx}$, $\rho_{xy}$
and $\rho_{yy}$.  Equation (\ref{eq:sup_inv_cond}) then imposes
constraints on the polynomial that take the
form of a system of linear equations for the
  coefficients. Their solution
give us the most
  general form of the invariant free density energy up to the order of
  the polynomial considered. For
  instance, at quadratic order the symmetry-restricted free energy
  density involves five independent terms:
\begin{equation}
f_2=a_1
(\Delta\rho)^2+a_2\det(\nabla\nabla\rho)+a_3\rho\Delta\rho+a_{4}|\nabla\rho|^2+a_{5}\rho^2,
\label{eq:sup_quad}
\end{equation}
The subscript $2$ for $f$ reminds the reader of the order
  of the polynomial considered. All terms here are scalars so there are no pseudoscalar
terms at quadratic order. Choosing
$a_1=1/2$, $a_3=k_0^2$, $a_{4}=0$ and $a_{5}=(k_0^2+r)/2$ reproduces
the quadratic contributions to the LB energy
density (\ref{eq:LBEnergy}). 

At third order we encounter one pseudoscalar term:
$$f_{3}=\nabla\rho\cdot(\nabla\nabla\rho)\cdot(\mathbf{n}\times\nabla\rho),$$
where $\mathbf{n}$ is the normal to the plane (co-incident with the
z-axis). This term is closely related to the Helfrich-Prost (HP) free energy density for chiral surfaces \cite{Helfrich}. There are two fourth-order pseudoscalar terms namely
$$f_4^a=\rho\nabla\rho\cdot(\nabla\nabla\rho)\cdot(\mathbf{n}\times\nabla\rho)$$
and
$$f_4^b=\nabla\rho\cdot(\nabla\nabla\rho)^2\cdot(\mathbf{n}\times\nabla\rho).$$

Each pseudoscalar density on a flat surfaces generates a corresponding pseudoscalar density on a curved surface that is obtained by replacing all partial derivatives by
  covariant derivatives and replacing $\mathbf{n}$ by the local unit normal
  to the surface. In Appendix B we show that the surface integrals of the covariant expressions for $f_{3}$ and $f_4^a$ over a spherical surface are zero. Our final expression for a chiral LB free energy thus only involves $f_4^b$

\begin{equation}
\begin{split}
  \mathcal{H}_\chi=&\int\Bigg(\frac{1}{2}\Big( (\Delta+k_o^2)\rho\Big)^2+\frac{r}{2}\thinspace\rho^2+\frac{u}{3}\thinspace\rho^3+\frac{v}{4}\thinspace\rho^4\\&+
\chi\nabla\rho\cdot(\nabla\nabla\rho)^2\cdot(\mathbf{n}\times\nabla\rho)\Bigg)\;dS.
  \label{eq:LBEnergy}
\end{split}
\end{equation}
where $\chi$ is a pseudoscalar that measures the strength of the chiral character of the interactions between the constituent units. 

It would seem reasonable to demand that the choice of the \textit{sign} of $\chi$ should be such that $\mathcal{H}_\chi$ to the free energy favors the odd $l$ sectors over the even $l$ sectors. However, we show in Appendix C that the integral
\begin{equation}
\begin{split}
&\int_{S^2}\left(\nabla Y_{l}^{m_1}\cdot(\nabla\nabla Y_{l}^{m_2})\cdot(\nabla\nabla Y_{l}^{m_3})\cdot(\mathbf{n}\times\nabla Y_{l}^{m_4})\right)\;dS=0
  \label{eq:LBEnergy}
\end{split}
\end{equation} 
is zero for \textit{any} set of $(m_1,m_2,m_3,m_4)$ and \textit{for any $l$}. The same is true if any of the spherical harmonics is replaced by its complex conjugate. The chiral term thus can not alter the minimization of the free energy when the minimization is restricted to only one given $l$ sector. This is an important result: \textit{in conventional single-irrep Landau theory, chirality would have no effect on orientational ordering.} This contradicts numerous observations on viral assembly and indicates that single-irrep Landau theory is unable to account for viral assembly,

\subsection{Chirality and the uniform state.}

We first need to demonstrate that a uniform state described by $\mathcal{H}_\chi$ is chiral. The free energy minimum in the uniform state is at a mean density modulation $\rho=0$ so the chirality of the uniform state cannot be determined by the properties of $\rho$ under inversion. Whether or not a state has broken chiral symmetry is determined by the response of the system to a probe that couples to density in a manner that is sensitive to the chirality of the state, such as polarization rotation. We will follow this route to determine whether the uniform state is chiral.
By repeating the previous arguments, it is easy to show that -- to lowest order in the density $\rho$ -- a local scalar chiral probe that couples to the density must have the form $\delta \mathcal{H}=-h_\chi\int f_4^b\;dS
$
Here, $h_\chi$ is an infinitesimal chiral scalar that measures the strength of the probe. The chirality $M_\chi$ of the system can be defined as $M_\chi=-(d\mathcal{F}/dh_\chi)_{h_\chi=0}$ where $\mathcal{F}$ is the free energy computed from the Hamiltonian $\mathcal{H}_\chi+ \delta \mathcal{H}$. It is easy to see that $M_\chi=<f_4^b>$ where $<..>$ indicates an average over the Boltzmann distribution $\mathcal{H}_\chi$. Calculating $M_\chi$ perturbatively in $\chi$, one finds that the lowest-order non-zero contribution to $M_\chi$ is proportional to $\beta\chi<(f_4^b)^2>$, where the thermal average is to be computed for $\mathcal{H}$ with $\chi=0$. Since $(f_4^b)^2$ is positive definite, this average is non-zero so the isotropic state is indeed chiral. Physically, though the average modulation density $\rho=0$, the thermal density fluctuations now are chiral.

\subsection{Chirality and even $l$ icosahedral states.}

We first consider the effect of the chiral term on the even $l$ icosahedral states that were stable, namely $l = 6, 10, 12$. To be specific, assume that $k_0 R$ is in the interval segment $6<k_0 R<7$ where $l=6$ icosahedral ordering takes place and that $t_6$ is slightly negative while the other $t_l$ still are positive. As before, let $\xi$ be the $l=6$ icosahedral order parameter computed within the $l=6$ sector. We saw that if only $l=6$ spherical harmonics are included then the chiral term has no effect. Allow coupling of the $l=6$ spherical harmonics to the spherical harmonics of the neighboring $l=5$ and $l=7$ segments, since $t_5$ and $t_7$ are the lowest $t_l$ after $t_6$. Note that neither $l=5$ nor $l=7$ supports icosahedral order. First, minimize the free energy in the $l=5+6$ composition space. Define the set $c_{-m}$ to be the set of eleven expansion coefficients of the $l=5$ sector with $m$ running from $-5$ to $+5$. The $l=5$ spherical harmonics will be mixed with the $l=6$ density if the variational free energy has terms linear  in of the $l=5$ terms. First consider the cubic and quartic nonlinearities of the achiral LB free energy. In a perturbation expansion, neither the cubic nor the quartic can produce terms that are linear in the $c_{m}^5$ because the $Y_{5}^{m}$ are odd under inversion while $\mathcal{Y}_h(6)$ is even. Next, if the chiral term is computed as a mixture of the $Y_{5}^{m}$ and the $Y_{6}^{m}$ then the only non-zero integrals were found to be composed of one factor of $\mathcal{Y}_h(6)$ and three factors of $Y_{5}^{m}$, which results in a polynomial that is the sum of terms of the form $\xi c_{m1}c_{m2}c_{m3}$. Since chiral mixing produces only \textit{cubic} terms in $c_m$, $\mathcal{Y}_h(6)$ type ordering does not entrain secondary ordering in the neighboring $l=5$ segment. The same is true for the the $l=7$ segment. Similar conclusions are arrived at for $l=10$ and $l=12$. The chiral term does not entrain non-icosahedral contributions as secondary order parameters.

Could the chiral term entrain other icosahedral spherical harmonics?  We find that the chiral term is non-zero for combinations that are third order in $\mathcal{Y}_h(6)$ but linear in $\mathcal{Y}_{h}(15)$. It follows that icosahedral ordering in the $l=6$ sector, with a density that is even under inversion, entrains $l=15$ icosahedral ordering, with a density that is odd under inversion. The same holds for the $l=10$ and $l=12$ states. The chiral term thus has an important consequence: it removes the objection against  primary icosahedral ordering in the $l=6, 10$ and $12$ segments because the chiral term generates secondary icosahedral contributions that now are odd under inversion. The resulting density is neither odd nor even under inversion, as is the case for the density of actula viral capsids.

\subsection{Chirality and the $l=15+16$ icosahedral state}

Now consider the effect of the chiral term on the isomeric pair of mixed $l=15+16$ icosahedral states. Assume again a main icosahedral density of the form \begin{equation}
  \rho(\zeta,\xi) = \zeta \mathcal{Y}_h(15) + \xi \mathcal{Y}_h (16),
  \label{eq:coupled_ansatz}
\end{equation}
and minimize $\mathcal{H}_{\chi}$ with respect to $\zeta$ and $\xi$. The resulting EL equations are
\begin{subequations}
  \begin{equation}
    (c_1+r)\xi + u_1 \xi^2 + u_2\zeta^2 + w_1 \xi^3+ w_2 \xi\zeta^2 + \chi(3a_1 \xi^2 \zeta+a_2\zeta^3)= 0,
    \label{xi}
  \end{equation}
  \begin{equation}
    (c_2+r)\zeta + u_3\xi\zeta + w_3 \zeta^3 + w_4\zeta \xi^2 +\chi(a_1 \xi^3 +3a_2\zeta^2\xi)= 0,
    \label{zeta}
  \end{equation}
  \label{zeta}
\end{subequations}
with both $a_2$ and $a_3$ positive. Recall that for $\chi=0$ there were two degenerate solutions ($D$ and $L$) related by $\zeta\rightarrow-\zeta$. The two terms proportional to $\chi$ in each equation lifts this degeneracy. To lowest order in $\chi$, the chiral term causes the free energy to shift by an amount $\chi\left(a_1\xi_0^3\zeta_0+a_2\xi_0\zeta_0^3\right)$ where ($\zeta_0,\xi_0$) denotes the $\chi=0$ solution. Since this term is odd in $\zeta_0$, the chiral term selects whether the $D$ or the $L$ isomer has the lower free energy. Because the chiral term breaks the symmetry between the isomers, there is no chiral symmetry breaking at the transition. The effect of the chiral term is thus very ``visible" in the $l=15+16$ superposition state.

\section{Examples and Conclusion.}
In Section I we discussed that the current interpretation of capsid
densities is based on comparing odd-$l$ icosahedral spherical
harmonics with measured capsid densities, as in Fig. 2. In the proposed chiral Landau-Brazovskii theory, even and odd spherical harmonics are mixed. In this section, we discuss the interpretation of chiral Landau-Brazovskii for a number of specific cases. Before doing that, we first must briefly review the Caspar-Klug (CK) classification of viral capsids \cite{caspar}.

\subsection{Caspar-Klug Construction}
The CK construction is based on the notion that identical capsid proteins should be distributed over an icosahedral capsid in a manner that minimizes local deformation of the proteins. Such deformations are the (unavoidable) consequence of the fact that not all sites can be symmetry-equivalent. The claim of CK theory is that such deformations are minimized by constructiing icosahedra in the manner shown in Fig. \ref{CK1}:

\begin{figure}[htb]
\centering
\includegraphics[width=0.45\textwidth]{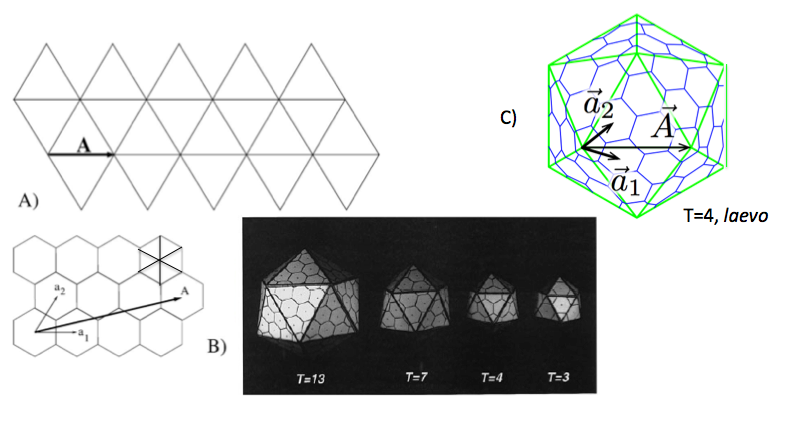}
\caption[Caspar-Klug Construction]{\label{CK1} Construction of icosahedra. A: A lattice vector of a two-dimensional hexagonal lattice is used to construct an equilateral triangle with vertices on lattice sites (from \cite{N1}). B: By gluing the edges of the folding template, icosahedra can be constructed (\cite{caspar}). C: A $T=7$ icosahedron with $h=2$ and $k=1$ is chiral.}\label{CK}
\end{figure}

Icosahedra are generated by cutting templates from a hexagonal sheet
composed of twenty adjacent equilateral triangles (see
Fig. \ref{CK1}A). The base of each triangle is a lattice vector
expressed as $\vec{A}(h,k)=h\hat{a}_1+k\hat{a}_2$. Here $\{h,k\}$ are
a pair of non-negative integers and $\hat{a}_{1,2}$ are a pair of
basis vectors of the hexagonal lattice (see Fig. \ref{CK1}A and
C). The icosahedron is constructed by pasting together adjacent
exposed edges of the template (see Fig. \ref{CK1}B). This construction
can be repeated for every pair of integers $h$ and
$k$. Fig. \ref{CK1}C shows the case of $h=3$ and $k=1$. The size of
the icosahedron is determined by the length of the base vector
$\vec{A}(h,k)$. It follows from simple geometry that twice the area
per triangle $|\vec{A}(h,k)|^2$ equals $T(h,k)=h^2+k^2+hk$. CK
icosahedra are fully characterized by this ``$T$ Number". As can be
verified from Fig. \ref{CK1}B, CK shells are composed of 12 pentagons and $10(T-1)$ hexagons. The smallest shell is $T=1$, composed of 12 pentagons, followed by the $T=3$ and $T=4$ shells shown in Fig. \ref{CK1}. The $T=1$, 3 and 4 icosahedra are invariant under inversion. The two larger shells shown in Fig. \ref{CK1} are $T=7$ and $T=13$. Each can be constructed in two separate, chirally asymmetric ways ($D$ and $L$) related by inversion. 
 
The CK icosahedra can be compared to the Bravais lattices of
solid-state physics: they are purely mathematical constructs. To
produce a physical capsid, capsid proteins must be assigned to the CK
icosahedra just as a ``basis" of molecules must be assigned to a
Bravais lattice to produce physical crystals. In the simplest case,
three proteins are placed on equivalent sites of each of the triangles
of Fig. \ref{CK}. Capsid proteins in general have no symmetry at all and are neither even nor odd under inversion. This \textit{extrinsic} source of chirality must be distinguished from the \textit{intrinsic} chirality of the $T=7$ and $T=13$ D and L CK shells. These would remain chiral even for (hypothetical) capsid proteins with an inversion center. 

Radial densities, measured by X-ray diffraction or electron microscopy and integrated across the thickness of the capsid, can be expanded in the icosahedral spherical harmonics $\mathcal{Y}_h(l)$.
Current Landau associates the primary icosahedral order parameter with a single $\mathcal{Y}_h(l)$. There are two concerns with this approach. The expansion of the density may involve a whole series of even and odd icosahedral spherical harmonics just as the expansion of the electron density of a crystal in general requires an extended series of wavevectors in the first Brillouin zone. It is not obvious that a density expansion of a viral capsid will be dominated by a single $l$. Next, densities measured by cryoEM or by X-ray diffraction need not be representative of the density of the capsid at the point of solidification. 

 In the next subsections we will consider (or reconsider) a number of specific examples of viral assembly from the viewpoint of the proposed theory and compare the predictions of the two Landau theories keeping these concerns in mind. We also revisit the question whether caspid solidification is a continuous or a first order transition. We shall see that determination of the primary order parameter for the Landau theory from measured densities may be ambiguous.

\subsubsection{Parvovirus.}
Our first two examples are $T=1$ viruses. The $T=1$ parvovirus,
already shown in Fig. 1, assembles reversibly from sixty monomers \cite{Tsao}. We saw  that $\mathcal{Y}_h(15)$ describes quite well the coarse-grained features of the measured density. Assume that the proteins in a (hypothetical) protein-RNA precursor state would remain \textit{monomeric}. In that case, one would expect that the ordering transition should be signaled by the development of an icosahedral density modulation $\rho(\Omega)$ with sixty maxima. Since $\mathcal{Y}_h(15)$ indeed has sixty maxima, there seems to be no ambiguity. Note however that $\mathcal{Y}_h(15)$ does not describe the fine-structure of the parvovirus capsid. Note, for example, the tiny 5-fold symmetric structure located at the five-fold symmetry location. Such contribution have to be described by \textit{secondary order parameters}  (i.e., multiplying $\mathcal{Y}_h(l)$ with $l$ larger than $15$) that are generated by the non-linear terms in the Landau energy. Though LB theory will generate such secondary order parameters, the correct expansion coefficients for any particular capsid  will require including additional non-linear terms in $\rho$ and its gradients to the Landau free energy. The corresponding coefficients would have to be obtained by fitting computed density modulations with measured ones.

The proposed Landau theory states that there must also be a contribution to the \textit{primary} order parameter that transforms as $\mathcal{Y}_h(16)$ under the symmetry operations of $I$. A  decomposition of the measured density of parvovirus might reveal the presence of such a contribution but it would be hard to distinguish that from a conventional secondary order parameter. The proposed Landau theory predict that the onset of icosahedral order either is a continuous transition or a discontinuous with, in a narrow stability range, a tetrahedral phase. Discovery of the tetrahedral phase would be strong evidence in favor of the proposed Landau theory.

\subsubsection{Picornavirus.}
\begin{figure}[htb]
\centering
\includegraphics[width=0.5\textwidth]{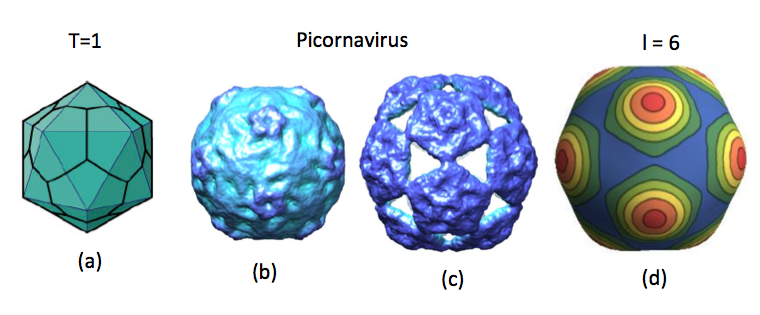}
\caption[Picorna Virus.]{(a) $T=1$ CK construction. (b) Solution structure of native picornavirus particles (the \textit{Equine rhinitis A} virus) and (c) the corresponding structure of the expanded particle (from ref. \cite{Bakker}).  (d) Interpretation as an $L=6$ orientational transition.} 
\label{picorna}
\end{figure}

Our second example concerns the \textit{picornaviruses}, a group of
animal viruses that includes the rhino and polioviruses (see Fig. \ref{picorna}b-c). Note that the capsid has a pronounced chiral character.
In solution, picorna virus capsid proteins are organized into stable
pentagons \cite{Zlotnick}. The pentagons are composed of five
asymmetric units, each of which is composed of three proteins. In
total, there are thus 15 proteins per pentagonal unit and 12 such
units per capsid for a total of 180 capsid proteins, which can be
indexed as a $T=3$ CK capsid. It is however known that the pentagons
survive as distinct units inside assembled picorna capsids because
picorna capsid can be swollen (by chemical treatment), causing the
pentagons to separately emerge while maintaining contact at their
vertices (see Fig. \ref{picorna}c). Treating the pentagons as separate entities leads to a $T=1$ assignment.

The choice of the primary order parameter leads to a dilemma. If it is
argued that 180 maxima should emerge at the point of solidification,
then this would be described by an icoosahedral spherical harmonic
with $l=27$, which has intrinsic chirality. If, on the other hand,
assembly is viewed as a $T=1$ process -- as in ref.\cite{Zlotnick} --
then the dominant value of $l$ should produce a density modulation
with 12 maxima, which corresponds to $\mathcal{Y}_h(6)$ (see Fig. \ref{picorna}d). In the second case, the chirality of the capsid should be due to odd-$l$ icosahedral spherical harmonics entrained as secondary order parameters by the $l=6$ primary order parameters (such as $\mathcal{Y}_h(15)$). In this view, it would \textit{not} be correct to extract the primary order parameter as the largest coefficient in an expansion of the density in icosahedral spherical harmonics. 

The two choices lead to different assembly thermodynamics. If the
primary order parameter is $l=6$ then the solidification process
should have the nature of a discontinuous transition. Because of
limitations, we have not been able to obtain the stability diagram for
$l=27$ but it is expected to be similar to that of Fig. 15 while in the current Landau theory it is described as a continuous transition.

\subsubsection{Cowpea Chlorotic Mottle Virus.} 
\begin{figure}[htb]
\centering
\includegraphics[width=0.5\textwidth]{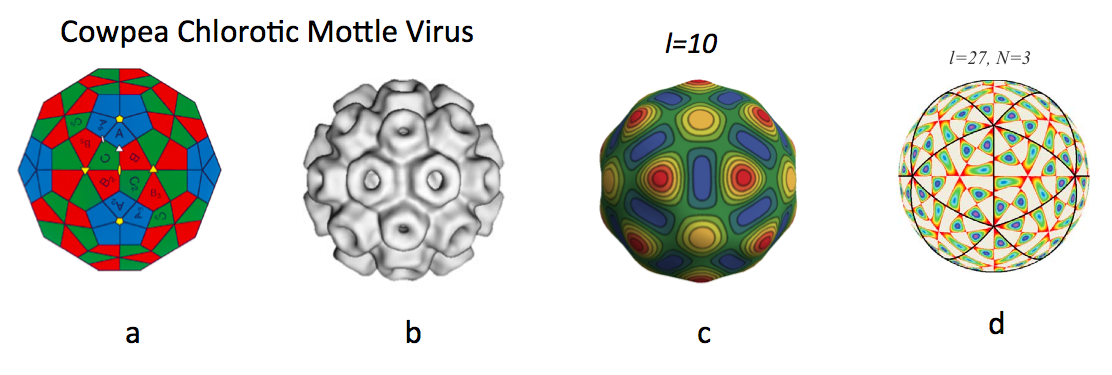}
\caption[Representation of a Caspar-Klug $T=3$ capsid]{\label{CK2} (a) A $T=3$ CK shell is divided up into twelve pentagons and twenty hexagons. Proteins are represented as kite-shaped units with 6 proteins per hexagons and 5 proteins per pentagon (b) Electron micrograph of the CCMV virus (from ref. \cite{liepold}). (c) Interpretation as an $l=10$ icosahedral spherical harmonic. (d) Interpretation as an $l=27$ icosaehdral spherical harmonic (from ref. \cite{lorman2008landau}).}
\label{CCMV}
\end{figure}

The Cowpea Chlorotic Mottle Virus (CCMV), a $T=3$ RNA plant virus, is
another important example because it is the one case for which a
precursor assembly state has directly been observed
\cite{Garmann}. Though CCMV capsid proteins in solution are dimers, at
least some fraction of the capsid proteins in the precursor state
appeared to be organized into \textit{capsomers}, i.e., pentamers and
hexamers of capsid proteins. Fig. \ref{CCMV}b shows an electron
micrograph of CCMV and compares it to a $T=3$ CK organizational
diagram (Fig. \ref{CCMV}a)
The capsomers are prominently visible in the assembled capsid. Other $T=3$ \textit{Bromoviridae} are organized in the same manner. Note that the micrograph does not have an obvious chiral character. Under inversion, the representation scheme only exchanges green and red colors. 

There are again two approaches possible. Assuming a CCMV capsid to be
assembled from 12 pentamers and 20 hexamers, the primary
$\rho(\Omega)$ of CCMV the capsomer organization should exhibit thirty
two maxima. This would correspond to $\mathcal{Y}_h(10)$, as shown in
Fig. \ref{CCMV}c. The assembly transition should be first order. Alternatively, if one assumes that assembly is signaled by the development of 180 maxima then $\mathcal{Y}_h(27)$ would again be the primary order parameter density. 

\subsubsection{Dengue Virus.}
\begin{figure}[htb]
\centering
\includegraphics[width=0.35\textwidth]{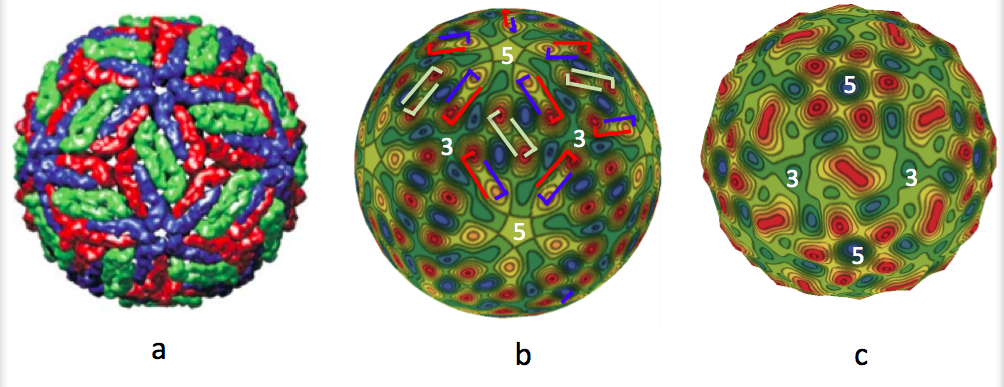}
\caption[Dengue virus.]{\label{CK2} (a) The Dengue virus is composed of 90 homodimers organized in a $T=3$ lattice. The three colors correspond to the three different symmetry environments for the capsid proteins (from ref. \cite{lorman2008landau}). (b) A $l=25+26$ icosahedral shell. High density (red/orange) correspond to the ends of the dimers. (c) A chiral superposition state of different icosahedral spherical harmonics with 90 maxima.}
\label{Dengue}
\end{figure}

Next, the Dengue virus is a $T=3$ RNA animal virus \cite{Dengue1},
composed of 180 subunits like CCMV but, unlike CCMV, there are no
compact hexamers or pentamers in the capsid organization and neither
do pentamers or hexamers form in solution (Fig. \ref{Dengue}a). 
The capsid is composed of elongated \textit{dimers}, outlined
schematically. Capsid proteins that border the 5-fold sites (shown as
blue) dimerize with the capsid proteins that border the 3-fold sites
(shown as red) The capsid proteins that occupy the 2-fold sites
(green) dimerize with themselves. As noted by in
ref.\cite{lorman2008landau}, the density of the Dengue virus capsid
matches nicely to $\mathcal{Y}_h(25)$ (see Fig. \ref{Dengue}b, density maxima correspond to the ends of the dimers). It was discussed in Section III that the ordering transition in the $l=25+26$ superposition space is first order and narrowly confined to a short interval of $k_0R$ values. 

But there is a problem: the dimers of the Dengue capsid virus are
stable in solution so assembly involves combining \textit{90} such
dimers into a capsid \cite{Dengue2}. This suggests that the primary
order parameter at the solidification transition should be an
icosahedral spherical harmonic with 90 maxima but plots of the the
$\mathcal{Y}_h(l)$ for $l$ less than or equal to $l=27$ do not reveal
any such. As shown in Fig. \ref{Dengue}c, it certainly is possible to construct an icosahedral state with 90 maxima by superposition. The coefficients $c^2(l)$ are in this case $0.014 (l=6), 0.15 (l=10), 0.04 (l=12), 0.03 (l=15), 0.14 (l=16) 0.26 (l=18), 0.26 (l=21), 0.02 (l=22), 0.08 (l=25), 0.01 (l=26)$ (normalized to add to one). It is an unsolved problem how an ordering transition signaled by the development of 90 maxima could fit in either the proposed or the existing Landau theory.

It might be argued that simply counting maxima is not appropriate because of the elongated nature of the dimers. Definition of the density of a dimer on the surface requires not just defining its center of mass but also its orientation. This means that $\mathcal{Y}_h(25)$ still could be the appropriate primary order parameter density.

\subsubsection{HK97}
A particularly interesting case is HK97 virus. This is a highly studied laevo $T=7$ bacteriophage virus \cite{conway1}. HK97 is prototypical for a family of related bacteriophage and herpes viruses with double-stranded DNA genomes (instead of the single-stranded RNA genomes discussed  so far). In most cases, the capsids in this familiy assemble on  top of a precursor spherical protein scaffold, which appears to offer a convenient realization for the application of theories of orientational phase transitions. However, HK97 is exceptional as it does not form on a precursor scaffold. Instead, the capsid proteins are linked by an interconnected net of $\Delta$ groups below the actual capsid \cite{Gertsman}. This net is believed to play the role of a scaffold. Both the $\Delta$ groups and the scaffold are removed as part of the maturation. 

\begin{figure}[htb]
\centering
\includegraphics[width=0.45\textwidth]{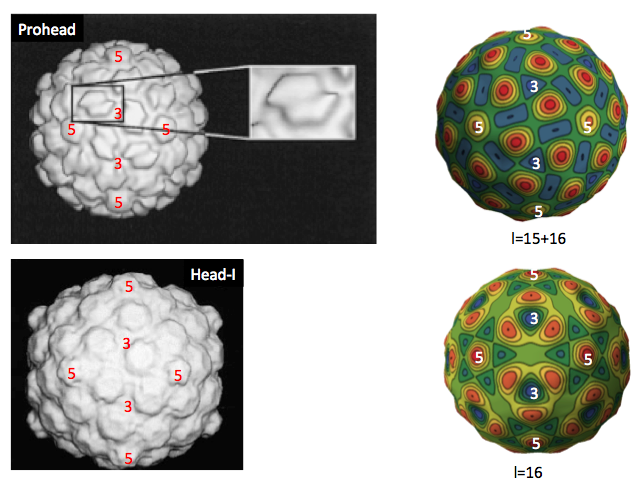}
\caption{Left top: Reconstruction of the first assembly stage of HK97 (Prohead 1). Three-fold and five-fold sites are indicated. The hexamers are highly sheared (from Ref.\cite{conway2}). Left bottom: Reconstruction of the first EI ``expanded" stage. The hexamers are symmetric. Right top: $l=15+16$ icosahedral capsid near the border with the $l=16$ sector. Right bottom: pure $l=16$}
\label{HK97}
\end{figure}

It has been established that, in solution, HK97 capsid proteins are
organized into stable pentamers and hexamers \cite{xie}. An HK97
capsid has 72 such capsomers. The $l=16$ icosahedral spherical
harmonic does have 72 capsomer and it provides a reasonable density
\textit{for a later stage} of the capsid maturation: the ``EI state"
(see Fig. \ref{HK97}, lower two panels)
The density of the EI state is practically achiral. This, however, is
not the state of the capsid immediately upon assembly. That is the
``Prohead" state, shown in the upper left panel of Fig. \ref{HK97}. The three hexamers adjacent to the three-fold sites are highly sheared \cite{Gertsman}, as shown in the figure, so the 3-fold sites have a pronounced chiral character. The 5-fold sites, however, are only weakly chiral. 

Landau theory provides an elegant account for the effect of symmetry on HK97 capsid development. A pure $l=16$ icosahedral spherical harmonic density modulation is unstable. By assembling the prohead from \textit{sheared} hexamers, which increases the chirality of the capsomer-capsomer interaction, a stable $l=15+16$ icosahedral state can form. Once the capsomers have established firm bonds with one another, the shear strain on the hexamers is released, which happens by the scissioning of the $\Delta$ groups from the capsid proteins, followed by injection of the DNA genome into the capsids. At this stage of the maturation, the capsid will not revert to the non-icosahedral stage. This importance of symmetry to control capsid assembly and maturation for the HK97 family is well appreciated \cite{Steven}.

According to the proposed Landau theory, ordering transition in the $15+16$ composition space with predominant $l=16$ character is expected to be first-order with a tetrahedral intermediate. It would be interesting to try and adjust the $k_0R$ parameter, for example through point mutations of the $\Delta$ sections, so as to reduce the shear strain of the hexamers. This should lead to the formation of shells with distorted \textit{octahedral} symmetry, which would be a very interesting discovery.

Experimental realization of any of these examples would be of great interest. Fluorescence thermal shift assay \cite{White} already makes it possible to probe whether an assembly transition is continuous or first order. When applied to CCMV, this method indicates that the assembly transition is first order \footnote{Tresset, G et al. Phys. Rev. Applied, accepted for publication.}. The decomposition of the density of a capsid into icosahedral spherical harmonics at the solidification transition of the precursor state has not yet been accomplished. However, we believe that the rapid development of cryoEM visualization methods of individual viruses will make such studies viable.

We conclude by reviewing limitations and possible extensions of the theory and then return to fundamental question concerning single-irrep Landau theory. The first limitation of the theory is its assumption of the presence of a rigid spherical scaffold that stabilizes the precursor state. Though there are instances for which this is a reasonable assumption, such as the Herpes Simplex virus, this premise is questionable for most cases. The observations on the CCMV precursor state \cite{Garmann} indicates that the RNA-protein condensate, though roughly spherical, fluctuates strongly. Including deformability of the surface is also important because of the relationship between Landau theory and the coarse-grained description of viral shells that is based on application of \textit{thin-shell elasticity theory} \cite{Lidmar}. This description, and its extensions, was found to account for shape changes of viral capsids, elastic deformation, and maturation, as well as the shapes of non-icosahedral shells. Though it can be shown that Landau theory accounts for one key ingredient -- ``in-plane" elastic deformation -- it leaves out the second key ingredient of ``out-of-plane" deformability.

A second limitation of the theory is its reliance on mean-field theory. Individual viruses are finite systems. Transitions of finite systems at finite temperature cannot have the thermodynamic singularities that are present in both the current and the proposed Landau theories of capsid assembly. Thermal fluctuations are expected to smear-out these singularities. 

An interesting extension of the theory would be to study its kinetics and the relation to the current nucleation-and-growth description of capsid assembly \cite{Zlotnick, Zandi}. Based on numerical simulations \cite{Hagan}, it is expected that when the binding affinity between different capsid proteins increases with respect to the binding affinity between capsid proteins and RNA then nucleation and growth will start to dominate over the collective scenario that is the focus of this paper. This is best described in a kinetic rather than an equilibrium context.

Finally, we return to the concerns about single-irrep Landau theory. We have seen that there are \textit{multiple} reasons that single-irrep Landau theroy cannot describe viral assembly. First, the only single-$l$ stable icosahedral states are $l=6$, 10, and 12. Second, the chiral term is zero when the state space is confined to a single value of $l$ so there would be no chiral capsids! Both claims are in clear conflict with observations on viral assembly. The original paper by Landau \cite{Landau1}, \footnote{Translated and reprinted in Landau L.D. Collected Papers (Nauka, Moscow, 1969), Vol. 1, pp. 234–252. Amusingly, Landau compares in this paper different irreps to different \textit{races}. In this context, the proposed theory could be viewed as a demonstration of the benefits of increased ethnic diversity.} states the single-irrep assumption without proof. In his textbook \cite{Landau2} it is argued that different irreps in general have different instability points so that multiple irreps only need to be considered at multi-critical points. The current case, in which  single irrep states are unstable while stability is regained in an extended composition space of multiple irreps, was not considered. The restriction to single irreps is not necessary and should be removed from Landau theory.

\section{Acknowledgments}
We would like to thank Alexander Grosberg for helpful discussions, the NSF for support under DMR Grant No. 1006128 and the Aspen Center for Physics for hosting a workshop on the physics of viral assembly. This paper is dedicated to the memory of our two friends and colleagues William Klug and Vladimir Lorman.

\newpage

\begin{appendix}

\section{The irrep basis of the icosahedral group and thermodynamic stability.}
In this appendix we replace the spherical harmonic basis functions with a new set associated with the irreducble representations of the symmetry group. We use it to discuss the stability of the famiiar case of the $l=6$ icosahedral spherical harmonic.

\begin{figure}[htbp]
\begin{center}
\includegraphics[width=2in]{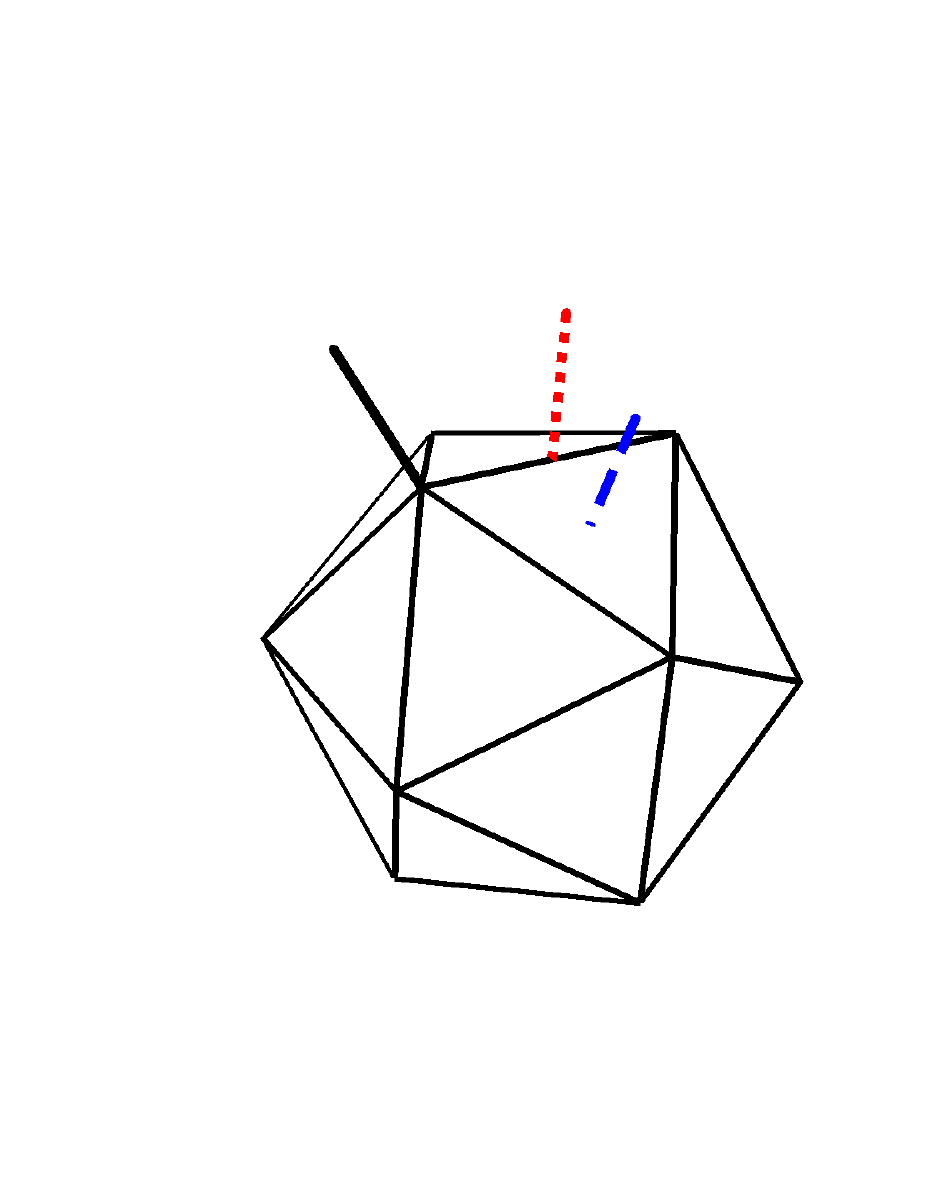}
\caption{The icosahedron, with examples of a five-fold, a three-fold, and a two-fold symmetry axis. The fivefold axis emerges from a vertex, the threefold axis from the center of a triangular face and the twofold axis from an edge.}
\label{fig:icosa1}
\end{center}
\end{figure}

The icosahedral group is composed of 60 rotations that map an
icosahedron into itself (see Fig. \ref{fig:icosa1}). 
They fall into five classes.  In the notation of Hamermesh
\cite{Hamermesh} these are: the identity $E$ with one member, two
rotations about a fivefold axis, $\mathcal{C}_5, \mathcal{C}_5^4$ with
12 members,  an additional two rotations about a fivefold axis,
$\mathcal{C}_5^2, \mathcal{C}_5^3$ with 12 members, the rotations
about twofold axes, $\mathcal{C}_2$, with 15 members, and the
rotations about threefold axes,  $\mathcal{C}_3, \mathcal{C}_3^2$,
with 20 members. The five-fold rotations are here about the axis
passing through two vertices on opposite sides of the icosahedron (see
Fig. \ref{fig:icosa1}); the three-fold rotations are about an axis that passes through the centers of two triangles on opposite side of the icosahedron and the two-fold rotations are through axes that pass through the center of edges on opposite sides of the polyhedron. The number of members of the five classes $1+12+12+15+20=60$ is the same as the number of elements of the group, as it should. For finite groups, the number of classes equals the number of irreducible representations (irreps) \cite{Hamermesh}. The five irreps of $I$ are listed in table 1 of the main text where $\chi_i^{(j)}$ where $\chi_i^{(j)}$ is the character of class $j$ for the $i^{\rm th}$ irrep.

Using this character table, one can project an appropriate mathematical object $\mathcal{O}$ onto a given irrep as follows:
\begin{equation}
\sum_{j=1}^{5} \chi_i^{(j)} \sum_{{ \bf R} \in C_j} { \bf R} \cdot \mathcal{O} \label{eq:icosa1}
\end{equation}
Here, {\bf R} is one of the rotation operations of $I$ while $C_j$ denotes the collection of symmetry operations of the $j$'th class while $i$ is the irrep in question. In the present case, the mathematical objects are the spherical harmonics belonging to a certain $l$. The action of the rotation operator {\bf R} on the spherical harmonic $Y_l^m(\widehat{r})$ produces $Y_l^m(\bf {R}(\widehat{r}))$. The rotated spherical harmonic can be expanded in the $2l+1$ spherical harmonics with the same $l$:
\begin{equation}
Y_l^m({ \bf R}(\widehat{r}))=\sum_{m'=-l}^{m'=+l}[D_{mm'}^l]^*Y_l^{m'}(\widehat{r})
\end{equation}
Here, $[D_{mm'}^l]^*$ is the complex conjugate of an element of the Wigner D-matrix, which is readily calculated \cite{Wigner}. For a given $l$, we can, in this way, generate $2l+1$ functions of $\widehat{r}$ that transform under the symmetry operations of the group as a particular irrep of that group.

\subsection{$l=6$}
As an example, consider the case of $l=6$. Operating on the 13 functions $Y_{6}^m(\widehat{r})$ with Eq. (\ref{eq:icosa1}) for the case of the one-dimensional irrep $A_g$ produces a 13 by 13 matrix:
\begin{equation}
\left(
\begin{array}{ccccccccccccc}
 0 & 0 & 0 & 0 & 0 & 0 & 0 & 0 & 0 & 0 & 0 & 0 & 0 \\
 0 & \frac{84}{5} & 0 & 0 & 0 & 0 & -\frac{12 \sqrt{77}}{5} & 0 & 0 & 0 & 0 &
   -\frac{84}{5} & 0 \\
 0 & 0 & 0 & 0 & 0 & 0 & 0 & 0 & 0 & 0 & 0 & 0 & 0 \\
 0 & 0 & 0 & 0 & 0 & 0 & 0 & 0 & 0 & 0 & 0 & 0 & 0 \\
 0 & 0 & 0 & 0 & 0 & 0 & 0 & 0 & 0 & 0 & 0 & 0 & 0 \\
 0 & 0 & 0 & 0 & 0 & 0 & 0 & 0 & 0 & 0 & 0 & 0 & 0 \\
 0 & -\frac{12 \sqrt{77}}{5} & 0 & 0 & 0 & 0 & \frac{132}{5} & 0 & 0 & 0 & 0 & \frac{12
   \sqrt{77}}{5} & 0 \\
 0 & 0 & 0 & 0 & 0 & 0 & 0 & 0 & 0 & 0 & 0 & 0 & 0 \\
 0 & 0 & 0 & 0 & 0 & 0 & 0 & 0 & 0 & 0 & 0 & 0 & 0 \\
 0 & 0 & 0 & 0 & 0 & 0 & 0 & 0 & 0 & 0 & 0 & 0 & 0 \\
 0 & 0 & 0 & 0 & 0 & 0 & 0 & 0 & 0 & 0 & 0 & 0 & 0 \\
 0 & -\frac{84}{5} & 0 & 0 & 0 & 0 & \frac{12 \sqrt{77}}{5} & 0 & 0 & 0 & 0 &
   \frac{84}{5} & 0 \\
 0 & 0 & 0 & 0 & 0 & 0 & 0 & 0 & 0 & 0 & 0 & 0 & 0 \\
\end{array}
\right) \label{eq:icosa2}
\end{equation}
One obtains results as the entries of a column vector generated by operating with this matrix on the column vector whose 13 entries are $Y_6^{m}( \theta, \phi)$. The column vectors that result from this operation have only three non-zero entries in the form of three function of $\widehat{r}$, the second, seventh and twelfth from the top. They are all  proportional to each other and to the icosahedral spherical harmonic $\mathcal{Y}_h(6)$. This means that there is exactly one combination of $l=6$ spherical harmonics that generates a density with full icosahedral symmetry and only one instance of the one-dimensional irrep of the icosahedral group that can be constructed by this method.

Performing similar operations for the four other irreps of the icosahedral group, we find that  one representative from each of three irreps can be constructed out of the $Y_6^m( \theta, \phi)$s. Those three irreps are the first three dimensional, the four dimensional and the five dimensional representation. Adding up dimensions, we have $1+3+4+5 =13$ dimensions, corresponding to 13 basis states, exactly as many as are provided by the 13 spherical harmonics for $l=6$. Thus, the basis states for the irreps provide an alternative basis for the analysis of the various properties of a ``crystallized'' $l=6$ state. The new basis set consists of normalized basis states of the irreducible representations, $\phi_i^{(j)}(\theta, \phi)$ where the subscript refers to the representation, and the superscript in parentheses refers to the basis within the representation. In the case of the one dimensional representation, corresponding to $ \phi_1( \theta, \phi) = \mathcal{Y}_h(6)$, the superscript is omitted. The normalization is
\begin{equation}
\int_0^{\pi} \int_0^{2 \pi} \phi_i^{(j)}( \theta, \phi)^2  \sin (\theta)d \phi \, d \theta = 1 \label{eq:normeq}
\end{equation}

It is now possible to recast the Landau Hamiltonian (\ref{eq:LBEnergy2}) in terms of these $l=6$ basis states. First assume that the only basis state participating is the one corresponding to perfect icosahedral symmetry, i.e. the identity irrep $A_g$. The effective Hamiltonian takes the usual form of a first-order transition
\begin{equation}
\frac{ t_6}{2}a_1^2 -\frac{10u}{969} \sqrt{\frac{143}{\pi }} a_1^3 +\frac{4719  v}{31280 \pi } a_1^4 \label{eq:icosa3}
\end{equation}
where $a_1$ is the amplitude of $ \phi_1( \theta, \phi) =\mathcal{Y}_h(6)$. When the  quadratic coefficient, $t_6$, is sufficiently small, a first order phase transition occurs to an ordered state, in which $a_1$ is non-zero; its sign is determined by the sign of the third order coefficient, $u$---positive for positive values of $u$ and negative for negative $u$ values. 

Determining the stability of this state entails allowing all basis states to participate in the Hamiltonian. This expansion is greatly simplified in the irrep-based expansion because it is \textit{entirely diagonal} so one can investigate the properties of each irrep separately. Furthermore all diagonal entries are the same for the basis states of a given irrep so can define effective temperatures $t_{6,i}$ for each irrep. 

For example, if $a_5^{(i)}$ is the amplitude of a basis state of the five-dimensional irrep $H_g$, then the quadratic term in the expansion of the effective Hamiltonian about the ordered state has the form
\begin{equation}
\begin{split}
&a_5^{(i) \, 2} \Big(\frac{t_6}{2} +3a_{1,s} \frac{u}{3} \int_0^{\pi}\int_0^{2 \pi} \phi_1(\theta, \phi) \phi_i^{(j) }( \theta, \phi)^2 \sin ( \theta) d \phi \, d \theta \\& + 6 a_{1,s}^2  \frac{v}{4} \int_0^{\pi}\int_0^{2 \pi} \phi_1(\theta, \phi)^2 \phi_i^{(j) }( \theta, \phi)^2 \sin ( \theta) d \phi \, d \theta \Big)  \label{eq:icosa3b}
\end{split}
\end{equation}
where only quadratic terms in the coefficient $a_5^{(i) }$ are included. The amplitude $a_{1,s}$ in (\ref{eq:icosa3b}) is the minimizing solution to Eq. (\ref{eq:icosa3}). The term inside the large brackets is half the effective temperature $t_{6,5}$ associated with $H_g$.

The integrals in (\ref{eq:icosa3b}) are all independent of superscript $(j)$. If the third order coefficient $u$ is zero and the quartic coefficient $v$ positive, then the term in parentheses turns is positive for negative $t_6$ and proportional to $|t_6|$. The precise relations are listed in Table \ref{table:icosa2} for the case $u=0$. The expressions in the table hold when $t_6<0$, corresponding to icosahedral ordering. 
\begin{table}[htp]
\caption{Values of the effective quadratic coupling for the four irreducible representations of the icosahedral group when $l=6$ The cubic coefficient, $u$ is equal to zero. The relationships hold when $t_6<0$}
\begin{center}
\begin{tabular}{|l|c|}
\hline
Irreducible representation & Quadratic coefficient, $t_{6,i}$ \\
\hline
\hline
One dimensional, $A_g$ & $2|t|_{6}$ \\
\hline
Three dimensional, $F_{1g}$ & 0 \\
\hline
Four dimensional, $G_g$ & $(42/2299) |t_6|$ \\
\hline
Five dimensional, $H_g$ & $(2492/11495) |t_6|$ \\
\hline
\end{tabular}
\end{center}
\label{table:icosa2}
\end{table}
In the case of the one dimensional representation, the quadratic coefficient governs fluctuations in the amplitude of the icosahedral order. The fact that there are three basis states with quadratic coefficient zero reflects the invariance of the energy of an ordered state with respect to overall rotations of the density.

\begin{figure}[htbp]
\begin{center}
\includegraphics[width=3in]{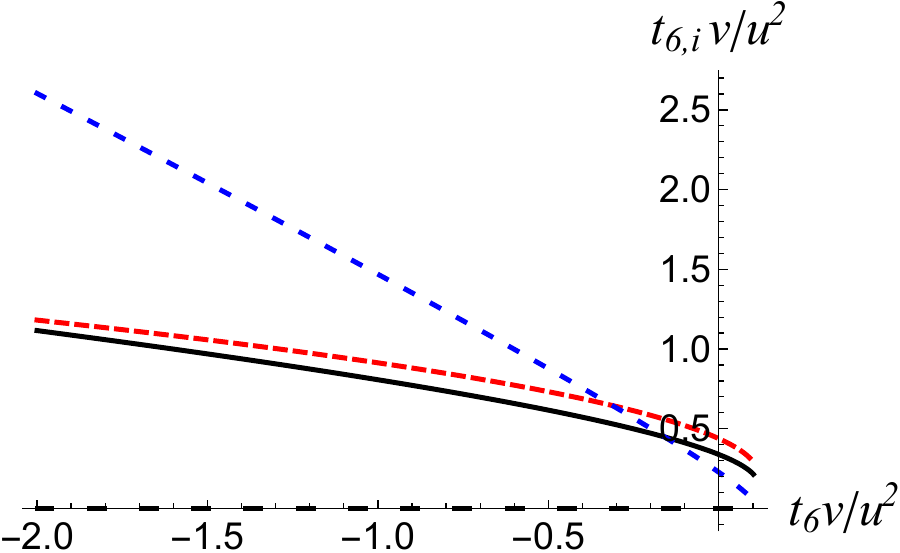}
\caption{The quadratic coefficients, $t_{6,i}$ when there is ordering for $l=6$. The four curves correspond to the one dimensional representation, (straight, blue dashed), the four dimensional representation (red, dashed), the five dimensional representation (black, solid) and the three dimensional representation (dashed, along the horizontal axis). }
\label{fig:l6stabilityplot}
\end{center}
\end{figure}

If $u$ is non-zero, then the expression in parentheses are more complicated. The quadratic coefficients, $t_{5,i}$, of the $a_6^{(i)}$'s are graphed in Fig. \ref{fig:l6stabilityplot}, expressed in terms of their ratio with respect to $u^2/v$.
The dashed line along the horizontal axis corresponds to the three dimensional representation and, again, reflects the fact that the density is insensitive to overall rotations of the sphere. Note that all these coefficients are either zero or positive. This means that the icosahedral state is locally stable.

\section{Chiral terms}

Without loss of generality, it is sufficient to consider the density
defined on a unit sphere $S^2$. Let
\begin{equation}
  I_3=\int_{S^2} \nabla\rho\cdot \nabla\nabla\rho\cdot(\mathbf{n}\times\nabla\rho)\;dS.
\end{equation}
which can be expressed as:
$$I_3=\int_{S^2}(\nabla^\alpha \rho) (\nabla_\alpha\nabla_\beta\rho) (\varepsilon^{\beta\nu}\nabla_\nu\rho)\;dS.$$
Integrating by parts,
\begin{equation}
  I_3=\int_{S^2}\nabla^\alpha\Big[\rho \varepsilon^{\beta\nu}\nabla_\nu\rho (\nabla_\alpha\nabla_\beta\rho)\Big]-\rho\nabla^\alpha\Big[\varepsilon^{\beta\nu}\nabla_\nu\rho (\nabla_\alpha\nabla_\beta\rho)\Big]\;dS.
\end{equation}
Since a sphere has no boundary, the first term is zero by divergence
theorem. Expanding the second term and using the definition of the
Laplace-Beltrami, $\nabla^\alpha\nabla_\alpha=:\Delta$,
$$I_3=-\int_{S^2}\rho
\Delta(\nabla_\beta\rho)(\varepsilon^{\beta\nu}\nabla_\nu\rho)+\rho(\nabla_\alpha\nabla_\beta\rho)\varepsilon^{\beta\nu}\nabla^\alpha\nabla_\nu\rho\;dS.$$
The second term involves the product of the anti-symmetric
$\varepsilon^{\beta\nu}$ and the symmetric
$\nabla_\alpha(\nabla_\beta\rho)\nabla^\alpha(\nabla_\nu\rho)$ and is
therefore zero. Using the following identity on the unit-sphere,
$$\Delta(\nabla_\beta)\rho=\nabla_\beta(\Delta\rho)+\nabla_\beta\rho,$$
we obtain,
$$I_3=-\int_{S^2}\rho
\varepsilon^{\beta\nu}\nabla_\beta(\Delta\rho)\nabla_\nu\rho + \rho
\varepsilon^{\beta\nu}\nabla_\nu\rho\nabla_\beta\rho\;dS.$$ Again, the second
term in this expression is zero as it involves the product of symmetric
and antisymmetric tensors. Applying the divergence theorem to the
first term,
$$I_3=\int_{S^2}(\Delta\rho)\varepsilon^{\beta\nu}\Big[\nabla_\beta\rho\nabla_\nu\rho+\rho\nabla_\beta\nabla_\nu\rho\Big]\;dS=0,$$
where the last equality follows from the observation that the
integrand is the product of symmetric and anti-symmetric tensors, thus
establishing the fact that the integral of the cubic chiral term over a spherical surface is zero.

Using this same method, it can be shown that one of the two quartic chiral terms is zero:
$$I_4^a=\int_{S^2}f_4^a\;dS=\int_{S^2}\rho \nabla\rho\cdot \nabla\nabla\rho\cdot(\mathbf{n}\times\nabla\rho)\;dS = 0.$$

The only non-trivial chiral term that we find is the quartic term,
\begin{equation}
  I_4^b=\int_{S^2}f_4^b\;dS = \int_{S^2} \nabla\rho\cdot (\nabla\nabla\rho)^2\cdot(\mathbf{n}\times\nabla\rho)\;dS.
  \label{eq:sup_quartic_chiral}
\end{equation}

\section{Chirality for a single irrep.}
In this section we show that the if the density $\rho$ is written in
terms of a single representation expansion, i.e., if
$$\rho=\sum_{m=-l}^lc_mY_l^m,$$ then the surface integral of the cubic chiral term is zero,
\begin{equation}
  I = \int_{S^2}\nabla\rho\cdot(\nabla\nabla\rho)^2\cdot(\mathbf{n}\times\nabla\rho)\;dS=0.
\end{equation}

To see this, rewrite $I$ in its coordinate representation,
$$I=\int_{S^2}(\nabla^\alpha\nabla_\beta\rho)(\nabla^\beta\rho)(\nabla_\alpha\nabla_\nu\rho)(\varepsilon^{\nu\gamma}\nabla_{\gamma}\rho)\;dS.$$
Using the product rule on the first two product terms of the integrand gives,
$$I=\frac{1}{2}\int_{S^2}\nabla^\alpha(|\nabla\rho|^2)(\nabla_\alpha\nabla_\nu\varepsilon^{\nu\gamma}\nabla_\gamma\rho)\;dS.$$
Using the divergence theorem, we obtain
$$2I
=\int_{S^2}\nabla^\alpha\Big[|\nabla\rho|^2(\nabla_\alpha\nabla_\nu\varepsilon^{\nu\gamma}\nabla_\gamma\rho)\Big] \hspace{.5in}$$
$$\hspace{1in}-|\nabla\rho|^2\nabla^\alpha[(\nabla_\alpha\nabla_\nu\rho)\varepsilon^{\nu\gamma}\nabla_\gamma\rho]\;dS.$$
Dropping the first surface term (since $S^2$ has no boundary) and expanding the gradient in the second term, we find
$$2I =
-\int_{S^2}|\nabla\rho|^2\Big[\Delta(\nabla_\nu\rho)\varepsilon^{\nu\gamma}\nabla_\gamma\rho
  +
  \varepsilon^{\nu\gamma}(\nabla_\alpha\nabla_\nu\rho)(\nabla^\alpha\nabla_\gamma\rho)\Big]\;dS.$$
The second term, being a product of a symmetric and an anti-symmetric
tensor, evaluates to zero. Using the identity, $\Delta(\nabla_\nu\rho)
= \nabla_\nu(\Delta\rho) + \nabla_\nu\rho$, for unit sphere, we can
rewrite the first term in previous integral as,
\begin{equation}
  I=-\frac{1}{2}\int_{S^2}|\nabla\rho|^2 \nabla(\Delta\rho)\times\nabla\rho\;dS,
  \label{eq:sup_chiral}
\end{equation}
having eliminated the term involving the product
$\varepsilon^{\nu\gamma}\nabla_\nu\rho\nabla_\gamma\rho$ as zero.Using the fact that
if $\rho=\sum_{m=-l}^lc_mY_l^m$ then
$\Delta\rho=-l(l+1)\rho$ it follows that, due to the cross product term
in the integrand, the integral (\ref{eq:sup_chiral}) evaluates to
zero.

This proof breaks down for an enlarged composition space containing multiple irreps. For
instance, if
$$\rho = \xi\mathcal{Y}_h(16)+\zeta\mathcal{Y}_h(15),$$ the
direct integration shows that $I$ is not zero.

\end{appendix}


\bibliography{mybib3}

\begin{thebibliography}{53}%
\makeatletter
\providecommand \@ifxundefined [1]{%
 \@ifx{#1\undefined}
}%
\providecommand \@ifnum [1]{%
 \ifnum #1\expandafter \@firstoftwo
 \else \expandafter \@secondoftwo
 \fi
}%
\providecommand \@ifx [1]{%
 \ifx #1\expandafter \@firstoftwo
 \else \expandafter \@secondoftwo
 \fi
}%
\providecommand \natexlab [1]{#1}%
\providecommand \enquote  [1]{``#1''}%
\providecommand \bibnamefont  [1]{#1}%
\providecommand \bibfnamefont [1]{#1}%
\providecommand \citenamefont [1]{#1}%
\providecommand \href@noop [0]{\@secondoftwo}%
\providecommand \href [0]{\begingroup \@sanitize@url \@href}%
\providecommand \@href[1]{\@@startlink{#1}\@@href}%
\providecommand \@@href[1]{\endgroup#1\@@endlink}%
\providecommand \@sanitize@url [0]{\catcode `\\12\catcode `\$12\catcode
  `\&12\catcode `\#12\catcode `\^12\catcode `\_12\catcode `\%12\relax}%
\providecommand \@@startlink[1]{}%
\providecommand \@@endlink[0]{}%
\providecommand \url  [0]{\begingroup\@sanitize@url \@url }%
\providecommand \@url [1]{\endgroup\@href {#1}{\urlprefix }}%
\providecommand \urlprefix  [0]{URL }%
\providecommand \Eprint [0]{\href }%
\providecommand \doibase [0]{http://dx.doi.org/}%
\providecommand \selectlanguage [0]{\@gobble}%
\providecommand \bibinfo  [0]{\@secondoftwo}%
\providecommand \bibfield  [0]{\@secondoftwo}%
\providecommand \translation [1]{[#1]}%
\providecommand \BibitemOpen [0]{}%
\providecommand \bibitemStop [0]{}%
\providecommand \bibitemNoStop [0]{.\EOS\space}%
\providecommand \EOS [0]{\spacefactor3000\relax}%
\providecommand \BibitemShut  [1]{\csname bibitem#1\endcsname}%
\let\auto@bib@innerbib\@empty
\bibitem [{\citenamefont {Onsager}(1949)}]{Onsager}%
  \BibitemOpen
  \bibfield  {author} {\bibinfo {author} {\bibfnamefont {L.}~\bibnamefont
  {Onsager}},\ }\href@noop {} {\bibfield  {journal} {\bibinfo  {journal} {Ann.
  NY Acad. Sci.}\ }\textbf {\bibinfo {volume} {51}},\ \bibinfo {pages} {627}
  (\bibinfo {year} {1949})}\BibitemShut {NoStop}%
\bibitem [{\citenamefont {Kats}\ \textit {et~al.}(1983)\citenamefont {Kats},
  \citenamefont {Lebedev},\ and\ \citenamefont {Muratov}}]{Kats}%
  \BibitemOpen
  \bibfield  {author} {\bibinfo {author} {\bibfnamefont {E.}~\bibnamefont
  {Kats}}, \bibinfo {author} {\bibfnamefont {V.}~\bibnamefont {Lebedev}}, \
  and\ \bibinfo {author} {\bibfnamefont {A.}~\bibnamefont {Muratov}},\
  }\href@noop {} {\bibfield  {journal} {\bibinfo  {journal} {Physics Reports}\
  }\textbf {\bibinfo {volume} {228}},\ \bibinfo {pages} {1} (\bibinfo {year}
  {1983})}\BibitemShut {NoStop}%
\bibitem [{\citenamefont {Steinhardt}\ \textit {et~al.}(1983)\citenamefont
  {Steinhardt}, \citenamefont {Nelson},\ and\ \citenamefont
  {Ronchetti}}]{Steinhardt}%
  \BibitemOpen
  \bibfield  {author} {\bibinfo {author} {\bibfnamefont {P.}~\bibnamefont
  {Steinhardt}}, \bibinfo {author} {\bibfnamefont {D.}~\bibnamefont {Nelson}},
  \ and\ \bibinfo {author} {\bibfnamefont {M.}~\bibnamefont {Ronchetti}},\
  }\href@noop {} {\bibfield  {journal} {\bibinfo  {journal} {Phys. Rev. B}\
  }\textbf {\bibinfo {volume} {28}},\ \bibinfo {pages} {784} (\bibinfo {year}
  {1983})}\BibitemShut {NoStop}%
\bibitem [{\citenamefont {Lorman}\ and\ \citenamefont
  {Rochal}(2008)}]{lorman2008landau}%
  \BibitemOpen
  \bibfield  {author} {\bibinfo {author} {\bibfnamefont {V.}~\bibnamefont
  {Lorman}}\ and\ \bibinfo {author} {\bibfnamefont {S.}~\bibnamefont
  {Rochal}},\ }\href@noop {} {\bibfield  {journal} {\bibinfo  {journal}
  {Physical Review B}\ }\textbf {\bibinfo {volume} {77}},\ \bibinfo {pages}
  {224109} (\bibinfo {year} {2008})}\BibitemShut {NoStop}%
\bibitem [{\citenamefont {Lorman}\ and\ \citenamefont
  {Rochal}(2007)}]{lorman2007density}%
  \BibitemOpen
  \bibfield  {author} {\bibinfo {author} {\bibfnamefont {V.}~\bibnamefont
  {Lorman}}\ and\ \bibinfo {author} {\bibfnamefont {S.}~\bibnamefont
  {Rochal}},\ }\href@noop {} {\bibfield  {journal} {\bibinfo  {journal}
  {Physical Review Letters}\ }\textbf {\bibinfo {volume} {98}},\ \bibinfo
  {pages} {185502} (\bibinfo {year} {2007})}\BibitemShut {NoStop}%
\bibitem [{\citenamefont {Cheng}\ \textit {et~al.}(1994)\citenamefont {Cheng},
  \citenamefont {Reddy}, \citenamefont {Olson}, \citenamefont {Fisher},
  \citenamefont {Baker},\ and\ \citenamefont {Johnson}}]{Johnson1994}%
  \BibitemOpen
  \bibfield  {author} {\bibinfo {author} {\bibfnamefont {R.~H.}\ \bibnamefont
  {Cheng}}, \bibinfo {author} {\bibfnamefont {V.~S.}\ \bibnamefont {Reddy}},
  \bibinfo {author} {\bibfnamefont {N.~H.}\ \bibnamefont {Olson}}, \bibinfo
  {author} {\bibfnamefont {A.~J.}\ \bibnamefont {Fisher}}, \bibinfo {author}
  {\bibfnamefont {T.~S.}\ \bibnamefont {Baker}}, \ and\ \bibinfo {author}
  {\bibfnamefont {J.~E.}\ \bibnamefont {Johnson}},\ }\href@noop {} {\bibfield
  {journal} {\bibinfo  {journal} {Structure}\ }\textbf {\bibinfo {volume}
  {2}},\ \bibinfo {pages} {271} (\bibinfo {year} {1994})}\BibitemShut {NoStop}%
\bibitem [{\citenamefont {Caspar}\ and\ \citenamefont {Klug}(1962)}]{caspar}%
  \BibitemOpen
  \bibfield  {author} {\bibinfo {author} {\bibfnamefont {D.~L.}\ \bibnamefont
  {Caspar}}\ and\ \bibinfo {author} {\bibfnamefont {A.}~\bibnamefont {Klug}},\
  }in\ \href@noop {} {\textit {\bibinfo {booktitle} {Cold Spring Harbor symposia
  on quantitative biology}}},\ Vol.~\bibinfo {volume} {27}\ (\bibinfo
  {organization} {Cold Spring Harbor Laboratory Press},\ \bibinfo {year}
  {1962})\ pp.\ \bibinfo {pages} {1--24}\BibitemShut {NoStop}%
\bibitem [{\citenamefont {Baker}\ \textit {et~al.}(1999)\citenamefont {Baker},
  \citenamefont {Olson},\ and\ \citenamefont {Fuller}}]{baker}%
  \BibitemOpen
  \bibfield  {author} {\bibinfo {author} {\bibfnamefont {T.}~\bibnamefont
  {Baker}}, \bibinfo {author} {\bibfnamefont {N.}~\bibnamefont {Olson}}, \ and\
  \bibinfo {author} {\bibfnamefont {S.}~\bibnamefont {Fuller}},\ }\href@noop {}
  {\bibfield  {journal} {\bibinfo  {journal} {Microbiology and Molecular
  Biology Reviews}\ }\textbf {\bibinfo {volume} {63}},\ \bibinfo {pages} {862}
  (\bibinfo {year} {1999})}\BibitemShut {NoStop}%
\bibitem [{Note1()}]{Note1}%
  \BibitemOpen
  \bibinfo {note} {Spheroidal retroviruses, like the Rous Sarcoma Virus \cite
  {Kingston}, and spheroidal bunyaviruses, like the Uukuniemi virus, are
  examples of spheroidal capsids that lack icosahedral symmetry \cite
  {Overby}.}\BibitemShut {Stop}%
\bibitem [{\citenamefont {Fraenkel-Conrat}\ and\ \citenamefont
  {Williams}(1955)}]{Fraenkel}%
  \BibitemOpen
  \bibfield  {author} {\bibinfo {author} {\bibfnamefont {H.}~\bibnamefont
  {Fraenkel-Conrat}}\ and\ \bibinfo {author} {\bibfnamefont {R.~C.}\
  \bibnamefont {Williams}},\ }\href@noop {} {\bibfield  {journal} {\bibinfo
  {journal} {Proc. Natl. Acad. Sci. U S A.}\ }\textbf {\bibinfo {volume}
  {41}},\ \bibinfo {pages} {690} (\bibinfo {year} {1955})}\BibitemShut
  {NoStop}%
\bibitem [{\citenamefont {Johnson}\ \textit {et~al.}(2005)\citenamefont
  {Johnson}, \citenamefont {Tang}, \citenamefont {Nyame}, \citenamefont
  {Willits},\ and\ \citenamefont {Zlotnick}}]{Zlotnick}%
  \BibitemOpen
  \bibfield  {author} {\bibinfo {author} {\bibfnamefont {J.}~\bibnamefont
  {Johnson}}, \bibinfo {author} {\bibfnamefont {J.}~\bibnamefont {Tang}},
  \bibinfo {author} {\bibfnamefont {Y.}~\bibnamefont {Nyame}}, \bibinfo
  {author} {\bibfnamefont {M.}~\bibnamefont {Willits}, \bibfnamefont
  {D.and~Young}}, \ and\ \bibinfo {author} {\bibfnamefont {A.}~\bibnamefont
  {Zlotnick}},\ }\href@noop {} {\bibfield  {journal} {\bibinfo  {journal} {Nano
  Lett.}\ }\textbf {\bibinfo {volume} {5}},\ \bibinfo {pages} {765} (\bibinfo
  {year} {2005})}\BibitemShut {NoStop}%
\bibitem [{\citenamefont {Zandi}\ \textit {et~al.}(2006)\citenamefont {Zandi},
  \citenamefont {van~der Schoot}, \citenamefont {Reguera}, \citenamefont
  {Kegel},\ and\ \citenamefont {Reiss}}]{Zandi2}%
  \BibitemOpen
  \bibfield  {author} {\bibinfo {author} {\bibfnamefont {R.}~\bibnamefont
  {Zandi}}, \bibinfo {author} {\bibfnamefont {P.}~\bibnamefont {van~der
  Schoot}}, \bibinfo {author} {\bibfnamefont {D.}~\bibnamefont {Reguera}},
  \bibinfo {author} {\bibfnamefont {W.}~\bibnamefont {Kegel}}, \ and\ \bibinfo
  {author} {\bibfnamefont {H.}~\bibnamefont {Reiss}},\ }\href@noop {}
  {\bibfield  {journal} {\bibinfo  {journal} {Biophys. J.}\ }\textbf {\bibinfo
  {volume} {90}},\ \bibinfo {pages} {1939} (\bibinfo {year}
  {2006})}\BibitemShut {NoStop}%
\bibitem [{\citenamefont {Hu}\ and\ \citenamefont
  {Shklovskii}(2007)}]{Shklovskii}%
  \BibitemOpen
  \bibfield  {author} {\bibinfo {author} {\bibfnamefont {T.}~\bibnamefont
  {Hu}}\ and\ \bibinfo {author} {\bibfnamefont {B.}~\bibnamefont
  {Shklovskii}},\ }\href@noop {} {\bibfield  {journal} {\bibinfo  {journal}
  {Phys. Rev. E}\ }\textbf {\bibinfo {volume} {75}},\ \bibinfo {pages} {051901}
  (\bibinfo {year} {2007})}\BibitemShut {NoStop}%
\bibitem [{\citenamefont {Garmann}\ \textit {et~al.}(2013)\citenamefont
  {Garmann}, \citenamefont {Comas-Garcia}, \citenamefont {Gopal}, \citenamefont
  {Knobler},\ and\ \citenamefont {Gelbart}}]{Garmann}%
  \BibitemOpen
  \bibfield  {author} {\bibinfo {author} {\bibfnamefont {R.}~\bibnamefont
  {Garmann}}, \bibinfo {author} {\bibfnamefont {M.}~\bibnamefont
  {Comas-Garcia}}, \bibinfo {author} {\bibfnamefont {A.}~\bibnamefont {Gopal}},
  \bibinfo {author} {\bibfnamefont {C.}~\bibnamefont {Knobler}}, \ and\
  \bibinfo {author} {\bibfnamefont {W.}~\bibnamefont {Gelbart}},\ }\href@noop
  {} {\bibfield  {journal} {\bibinfo  {journal} {J. Mol. Biol.}\ }\textbf
  {\bibinfo {volume} {426}},\ \bibinfo {pages} {1050} (\bibinfo {year}
  {2013})}\BibitemShut {NoStop}%
\bibitem [{\citenamefont {Klein}(1913)}]{Klein}%
  \BibitemOpen
  \bibfield  {author} {\bibinfo {author} {\bibfnamefont {F.}~\bibnamefont
  {Klein}},\ }\href@noop {} {\textit {\bibinfo {title} {Lectures on the
  icosahedron.}}}\ (\bibinfo  {publisher} {Dover Phoenix Editions},\ \bibinfo
  {year} {1913})\BibitemShut {NoStop}%
\bibitem [{\citenamefont {Golubitsky}\ \textit {et~al.}(2012)\citenamefont
  {Golubitsky}, \citenamefont {Stewart} \textit
  {et~al.}}]{golubitsky2012singularities}%
  \BibitemOpen
  \bibfield  {author} {\bibinfo {author} {\bibfnamefont {M.}~\bibnamefont
  {Golubitsky}}, \bibinfo {author} {\bibfnamefont {I.}~\bibnamefont {Stewart}},
   \textit {et~al.},\ }\href@noop {} {\textit {\bibinfo {title} {Singularities and
  groups in bifurcation theory}}},\ Vol.~\bibinfo {volume} {2}\ (\bibinfo
  {publisher} {Springer Science \& Business Media},\ \bibinfo {year}
  {2012})\BibitemShut {NoStop}%
\bibitem [{\citenamefont {Jaric}(1985)}]{Jaric}%
  \BibitemOpen
  \bibfield  {author} {\bibinfo {author} {\bibfnamefont {M.}~\bibnamefont
  {Jaric}},\ }\href@noop {} {\bibfield  {journal} {\bibinfo  {journal} {Phys.
  Rev. Lett.}\ }\textbf {\bibinfo {volume} {55}},\ \bibinfo {pages} {607}
  (\bibinfo {year} {1985})}\BibitemShut {NoStop}%
\bibitem [{\citenamefont {Tsao}\ and\ \citenamefont {et~al.}(1991)}]{Tsao}%
  \BibitemOpen
  \bibfield  {author} {\bibinfo {author} {\bibfnamefont {T.}~\bibnamefont
  {Tsao}}\ and\ \bibinfo {author} {\bibnamefont {et~al.}},\ }\href@noop {}
  {\bibfield  {journal} {\bibinfo  {journal} {Science}\ }\textbf {\bibinfo
  {volume} {251}},\ \bibinfo {pages} {1456} (\bibinfo {year}
  {1991})}\BibitemShut {NoStop}%
\bibitem [{Note2()}]{Note2}%
  \BibitemOpen
  \bibinfo {note} {All results in this paper are restricted to mean-field
  theory, and may be subject to corrections due to thermal
  fluctuations}\BibitemShut {NoStop}%
\bibitem [{\citenamefont {Collins}\ \textit {et~al.}(1997)\citenamefont
  {Collins}, \citenamefont {Sheldrake},\ and\ \citenamefont
  {Crosby}}]{collins}%
  \BibitemOpen
  \bibfield  {author} {\bibinfo {author} {\bibfnamefont {A.~N.}\ \bibnamefont
  {Collins}}, \bibinfo {author} {\bibfnamefont {G.}~\bibnamefont {Sheldrake}},
  \ and\ \bibinfo {author} {\bibfnamefont {J.}~\bibnamefont {Crosby}},\
  }\href@noop {} {\ \textbf {\bibinfo {volume} {2, pg. 151}},\ \bibinfo {pages}
  {151} (\bibinfo {year} {1997})}\BibitemShut {NoStop}%
\bibitem [{\citenamefont {Dharmavaram}\ \textit {et~al.}(2016)\citenamefont
  {Dharmavaram}, \citenamefont {Xie}, \citenamefont {Klug}, \citenamefont
  {Rudnick},\ and\ \citenamefont {Bruinsma}}]{Sanjay}%
  \BibitemOpen
  \bibfield  {author} {\bibinfo {author} {\bibfnamefont {S.}~\bibnamefont
  {Dharmavaram}}, \bibinfo {author} {\bibfnamefont {F.}~\bibnamefont {Xie}},
  \bibinfo {author} {\bibfnamefont {W.}~\bibnamefont {Klug}}, \bibinfo {author}
  {\bibfnamefont {J.}~\bibnamefont {Rudnick}}, \ and\ \bibinfo {author}
  {\bibfnamefont {R.}~\bibnamefont {Bruinsma}},\ }\href
  {http://stacks.iop.org/0295-5075/116/i=2/a=26002} {\bibfield  {journal}
  {\bibinfo  {journal} {EPL (Europhysics Letters)}\ }\textbf {\bibinfo {volume}
  {116}},\ \bibinfo {pages} {26002} (\bibinfo {year} {2016})}\BibitemShut
  {NoStop}%
\bibitem [{Note3()}]{Note3}%
  \BibitemOpen
  \bibinfo {note} {Landau Theory and the Emergence of Chirality in Viral
  Capsids. arXiv preprint arXiv:1606.02769 (2016).}\BibitemShut {Stop}%
\bibitem [{Note4()}]{Note4}%
  \BibitemOpen
  \bibinfo {note} {L. D. Landau and E. M. Lifshitz, "Statistical Physics,"
  chap. 14, section 136, Pergamon Press, Addison-Wesley Publishing Company,
  Inc., Reading, Mass., 1958.}\BibitemShut {Stop}%
\bibitem [{\citenamefont {Brazovskii}(1975)}]{brazovskii1975phase}%
  \BibitemOpen
  \bibfield  {author} {\bibinfo {author} {\bibfnamefont {S.}~\bibnamefont
  {Brazovskii}},\ }\href@noop {} {\bibfield  {journal} {\bibinfo  {journal}
  {Zh. Eksp. Teor. Fiz}\ }\textbf {\bibinfo {volume} {68}},\ \bibinfo {pages}
  {175} (\bibinfo {year} {1975})}\BibitemShut {NoStop}%
\bibitem [{\citenamefont {Pezzutti}\ \textit {et~al.}(2011)\citenamefont
  {Pezzutti}, \citenamefont {Vega},\ and\ \citenamefont {Villar}}]{Pezzutti}%
  \BibitemOpen
  \bibfield  {author} {\bibinfo {author} {\bibfnamefont {A.}~\bibnamefont
  {Pezzutti}}, \bibinfo {author} {\bibfnamefont {D.}~\bibnamefont {Vega}}, \
  and\ \bibinfo {author} {\bibfnamefont {M.}~\bibnamefont {Villar}},\
  }\href@noop {} {\bibfield  {journal} {\bibinfo  {journal} {Phil. Trans. R.
  Soc. A}\ }\textbf {\bibinfo {volume} {369}},\ \bibinfo {pages} {335}
  (\bibinfo {year} {2011})}\BibitemShut {NoStop}%
\bibitem [{\citenamefont {Sattinger}(1978)}]{sattinger1978bifurcation}%
  \BibitemOpen
  \bibfield  {author} {\bibinfo {author} {\bibfnamefont {D.}~\bibnamefont
  {Sattinger}},\ }\href@noop {} {\bibfield  {journal} {\bibinfo  {journal}
  {Journal of Mathematical Physics}\ }\textbf {\bibinfo {volume} {19}},\
  \bibinfo {pages} {1720} (\bibinfo {year} {1978})}\BibitemShut {NoStop}%
\bibitem [{\citenamefont {Wigner}(1959)}]{Wigner}%
  \BibitemOpen
  \bibfield  {author} {\bibinfo {author} {\bibfnamefont {E.~P.}\ \bibnamefont
  {Wigner}},\ }\href@noop {} {\textit {\bibinfo {title} {Group theory and its
  application to the quantum mechanics of atomic spectra}}},\ \bibinfo
  {edition} {expanded and improved}\ ed.,\ Pure and applied physics,\ (\bibinfo
   {publisher} {Academic Press},\ \bibinfo {address} {New York,},\ \bibinfo
  {year} {1959})\ p.\ \bibinfo {pages} {372 p.}\BibitemShut {Stop}%
\bibitem [{Note5()}]{Note5}%
  \BibitemOpen
  \bibinfo {note} {See Appendix A}\BibitemShut {NoStop}%
\bibitem [{Note6()}]{Note6}%
  \BibitemOpen
  \bibinfo {note} {$t_{16}=((k_0R)^2-16\times 17)^2+r$,
  $t_{15}=((k_0R)^2-15\times 16)^2+r$, $a_1/16=-80.4$, $a_2/16=-3084.1$,
  $u_1=0.13494 u$, $u_2=0.234946 u$, $v_1=1.04204 w$, $v2=0.681228 w$,
  $u_3=0.623577 u$, $v_3=0.789107 w$, and $v_4=0.904033 w$}\BibitemShut
  {NoStop}%
\bibitem [{\citenamefont {Hoyle}(2004)}]{Hoyle}%
  \BibitemOpen
  \bibfield  {author} {\bibinfo {author} {\bibfnamefont {R.}~\bibnamefont
  {Hoyle}},\ }\href@noop {} {\bibfield  {journal} {\bibinfo  {journal} {Physica
  D}\ }\textbf {\bibinfo {volume} {191}},\ \bibinfo {pages} {261} (\bibinfo
  {year} {2004})}\BibitemShut {NoStop}%
\bibitem [{\citenamefont {de~Gennes}\ and\ \citenamefont
  {Prost}(1993)}]{deGennes}%
  \BibitemOpen
  \bibfield  {author} {\bibinfo {author} {\bibfnamefont {P.}~\bibnamefont
  {de~Gennes}}\ and\ \bibinfo {author} {\bibfnamefont {J.}~\bibnamefont
  {Prost}},\ }\href@noop {} {\textit {\bibinfo {title} {Physics of Liquid
  Crystals}}}\ (\bibinfo  {publisher} {Oxford},\ \bibinfo {year}
  {1993})\BibitemShut {NoStop}%
\bibitem [{\citenamefont {Helfrich}\ and\ \citenamefont
  {Prost}(1988)}]{Helfrich}%
  \BibitemOpen
  \bibfield  {author} {\bibinfo {author} {\bibfnamefont {W.}~\bibnamefont
  {Helfrich}}\ and\ \bibinfo {author} {\bibfnamefont {J.}~\bibnamefont
  {Prost}},\ }\href@noop {} {\bibfield  {journal} {\bibinfo  {journal} {Phys.
  Rev. A.}\ }\textbf {\bibinfo {volume} {38}},\ \bibinfo {pages} {3065}
  (\bibinfo {year} {1988})}\BibitemShut {NoStop}%
\bibitem [{\citenamefont {Nguyen}\ \textit {et~al.}(2005)\citenamefont {Nguyen},
  \citenamefont {Bruinsma},\ and\ \citenamefont {Gelbart}}]{N1}%
  \BibitemOpen
  \bibfield  {author} {\bibinfo {author} {\bibfnamefont {T.}~\bibnamefont
  {Nguyen}}, \bibinfo {author} {\bibfnamefont {R.}~\bibnamefont {Bruinsma}}, \
  and\ \bibinfo {author} {\bibfnamefont {W.}~\bibnamefont {Gelbart}},\
  }\href@noop {} {\bibfield  {journal} {\bibinfo  {journal} {Physical Review
  E}\ }\textbf {\bibinfo {volume} {72}},\ \bibinfo {pages} {051923} (\bibinfo
  {year} {2005})}\BibitemShut {NoStop}%
\bibitem [{\citenamefont {Bakker}\ \textit {et~al.}(2014)\citenamefont {Bakker},
  \citenamefont {Groppelli}, \citenamefont {Pearson}, \citenamefont {Stockley},
  \citenamefont {Rowlands},\ and\ \citenamefont {Ranson}}]{Bakker}%
  \BibitemOpen
  \bibfield  {author} {\bibinfo {author} {\bibfnamefont {S.~E.}\ \bibnamefont
  {Bakker}}, \bibinfo {author} {\bibfnamefont {E.}~\bibnamefont {Groppelli}},
  \bibinfo {author} {\bibfnamefont {A.~R.}\ \bibnamefont {Pearson}}, \bibinfo
  {author} {\bibfnamefont {P.~G.}\ \bibnamefont {Stockley}}, \bibinfo {author}
  {\bibfnamefont {D.~J.}\ \bibnamefont {Rowlands}}, \ and\ \bibinfo {author}
  {\bibfnamefont {N.~A.}\ \bibnamefont {Ranson}},\ }\href@noop {} {\bibfield
  {journal} {\bibinfo  {journal} {Journal of virology}\ }\textbf {\bibinfo
  {volume} {88}},\ \bibinfo {pages} {6093} (\bibinfo {year}
  {2014})}\BibitemShut {NoStop}%
\bibitem [{\citenamefont {Liepold}\ \textit {et~al.}(2005)\citenamefont
  {Liepold}, \citenamefont {Revis}, \citenamefont {Allen}, \citenamefont
  {Oltrogge}, \citenamefont {Young},\ and\ \citenamefont {Douglas}}]{liepold}%
  \BibitemOpen
  \bibfield  {author} {\bibinfo {author} {\bibfnamefont {L.~O.}\ \bibnamefont
  {Liepold}}, \bibinfo {author} {\bibfnamefont {J.}~\bibnamefont {Revis}},
  \bibinfo {author} {\bibfnamefont {M.}~\bibnamefont {Allen}}, \bibinfo
  {author} {\bibfnamefont {L.}~\bibnamefont {Oltrogge}}, \bibinfo {author}
  {\bibfnamefont {M.}~\bibnamefont {Young}}, \ and\ \bibinfo {author}
  {\bibfnamefont {T.}~\bibnamefont {Douglas}},\ }\href@noop {} {\bibfield
  {journal} {\bibinfo  {journal} {Physical biology}\ }\textbf {\bibinfo
  {volume} {2}},\ \bibinfo {pages} {S166} (\bibinfo {year} {2005})}\BibitemShut
  {NoStop}%
\bibitem [{\citenamefont {Kuhn}\ \textit {et~al.}(2002)\citenamefont {Kuhn},
  \citenamefont {Zhang}, \citenamefont {Rossmann}, \citenamefont {Pletnev},
  \citenamefont {Corver}, \citenamefont {Lenches}, \citenamefont {Jones},
  \citenamefont {Mukhopadhyay}, \citenamefont {Chipman}, \citenamefont
  {Strauss} \textit {et~al.}}]{Dengue1}%
  \BibitemOpen
  \bibfield  {author} {\bibinfo {author} {\bibfnamefont {R.~J.}\ \bibnamefont
  {Kuhn}}, \bibinfo {author} {\bibfnamefont {W.}~\bibnamefont {Zhang}},
  \bibinfo {author} {\bibfnamefont {M.~G.}\ \bibnamefont {Rossmann}}, \bibinfo
  {author} {\bibfnamefont {S.~V.}\ \bibnamefont {Pletnev}}, \bibinfo {author}
  {\bibfnamefont {J.}~\bibnamefont {Corver}}, \bibinfo {author} {\bibfnamefont
  {E.}~\bibnamefont {Lenches}}, \bibinfo {author} {\bibfnamefont {C.~T.}\
  \bibnamefont {Jones}}, \bibinfo {author} {\bibfnamefont {S.}~\bibnamefont
  {Mukhopadhyay}}, \bibinfo {author} {\bibfnamefont {P.~R.}\ \bibnamefont
  {Chipman}}, \bibinfo {author} {\bibfnamefont {E.~G.}\ \bibnamefont
  {Strauss}},  \textit {et~al.},\ }\href@noop {} {\bibfield  {journal} {\bibinfo
  {journal} {Cell}\ }\textbf {\bibinfo {volume} {108}},\ \bibinfo {pages} {717}
  (\bibinfo {year} {2002})}\BibitemShut {NoStop}%
\bibitem [{\citenamefont {L{\'o}pez}\ \textit {et~al.}(2009)\citenamefont
  {L{\'o}pez}, \citenamefont {Gil}, \citenamefont {Lazo}, \citenamefont
  {Men{\'e}ndez}, \citenamefont {Marcos}, \citenamefont {S{\'a}nchez},
  \citenamefont {Vald{\'e}s}, \citenamefont {Falc{\'o}n}, \citenamefont
  {Mar{\'\i}a}, \citenamefont {M{\'a}rquez} \textit {et~al.}}]{Dengue2}%
  \BibitemOpen
  \bibfield  {author} {\bibinfo {author} {\bibfnamefont {C.}~\bibnamefont
  {L{\'o}pez}}, \bibinfo {author} {\bibfnamefont {L.}~\bibnamefont {Gil}},
  \bibinfo {author} {\bibfnamefont {L.}~\bibnamefont {Lazo}}, \bibinfo {author}
  {\bibfnamefont {I.}~\bibnamefont {Men{\'e}ndez}}, \bibinfo {author}
  {\bibfnamefont {E.}~\bibnamefont {Marcos}}, \bibinfo {author} {\bibfnamefont
  {J.}~\bibnamefont {S{\'a}nchez}}, \bibinfo {author} {\bibfnamefont
  {I.}~\bibnamefont {Vald{\'e}s}}, \bibinfo {author} {\bibfnamefont
  {V.}~\bibnamefont {Falc{\'o}n}}, \bibinfo {author} {\bibfnamefont
  {C.}~\bibnamefont {Mar{\'\i}a}}, \bibinfo {author} {\bibfnamefont
  {G.}~\bibnamefont {M{\'a}rquez}},  \textit {et~al.},\ }\href@noop {} {\bibfield
   {journal} {\bibinfo  {journal} {Archives of virology}\ }\textbf {\bibinfo
  {volume} {154}},\ \bibinfo {pages} {695} (\bibinfo {year}
  {2009})}\BibitemShut {NoStop}%
\bibitem [{\citenamefont {Conway}\ \textit {et~al.}(2001)\citenamefont {Conway},
  \citenamefont {Wikoff}, \citenamefont {Cheng}, \citenamefont {Duda},
  \citenamefont {Hendrix}, \citenamefont {Johnson},\ and\ \citenamefont
  {Steven}}]{conway1}%
  \BibitemOpen
  \bibfield  {author} {\bibinfo {author} {\bibfnamefont {J.}~\bibnamefont
  {Conway}}, \bibinfo {author} {\bibfnamefont {W.}~\bibnamefont {Wikoff}},
  \bibinfo {author} {\bibfnamefont {N.}~\bibnamefont {Cheng}}, \bibinfo
  {author} {\bibfnamefont {R.}~\bibnamefont {Duda}}, \bibinfo {author}
  {\bibfnamefont {R.}~\bibnamefont {Hendrix}}, \bibinfo {author} {\bibfnamefont
  {J.}~\bibnamefont {Johnson}}, \ and\ \bibinfo {author} {\bibfnamefont
  {A.}~\bibnamefont {Steven}},\ }\href@noop {} {\bibfield  {journal} {\bibinfo
  {journal} {Science}\ }\textbf {\bibinfo {volume} {292}},\ \bibinfo {pages}
  {744} (\bibinfo {year} {2001})}\BibitemShut {NoStop}%
\bibitem [{\citenamefont {Gertsman}\ \textit {et~al.}(2009)\citenamefont
  {Gertsman}, \citenamefont {Gan}, \citenamefont {Guttman}, \citenamefont
  {Lee}, \citenamefont {Speir}, \citenamefont {Duda}, \citenamefont {Hendrix},
  \citenamefont {Komives},\ and\ \citenamefont {Johnson}}]{Gertsman}%
  \BibitemOpen
  \bibfield  {author} {\bibinfo {author} {\bibfnamefont {I.}~\bibnamefont
  {Gertsman}}, \bibinfo {author} {\bibfnamefont {L.}~\bibnamefont {Gan}},
  \bibinfo {author} {\bibfnamefont {M.}~\bibnamefont {Guttman}}, \bibinfo
  {author} {\bibfnamefont {K.}~\bibnamefont {Lee}}, \bibinfo {author}
  {\bibfnamefont {J.}~\bibnamefont {Speir}}, \bibinfo {author} {\bibfnamefont
  {R.}~\bibnamefont {Duda}}, \bibinfo {author} {\bibfnamefont {R.}~\bibnamefont
  {Hendrix}}, \bibinfo {author} {\bibfnamefont {E.}~\bibnamefont {Komives}}, \
  and\ \bibinfo {author} {\bibfnamefont {J.}~\bibnamefont {Johnson}},\
  }\href@noop {} {\bibfield  {journal} {\bibinfo  {journal} {Nature}\ }\textbf
  {\bibinfo {volume} {458}},\ \bibinfo {pages} {646} (\bibinfo {year}
  {2009})}\BibitemShut {NoStop}%
\bibitem [{\citenamefont {Xie}\ and\ \citenamefont {Hendrix}(1995)}]{xie}%
  \BibitemOpen
  \bibfield  {author} {\bibinfo {author} {\bibfnamefont {Z.}~\bibnamefont
  {Xie}}\ and\ \bibinfo {author} {\bibfnamefont {R.}~\bibnamefont {Hendrix}},\
  }\href@noop {} {\bibfield  {journal} {\bibinfo  {journal} {J. Mol. Biol.}\
  }\textbf {\bibinfo {volume} {253}},\ \bibinfo {pages} {74} (\bibinfo {year}
  {1995})}\BibitemShut {NoStop}%
\bibitem [{\citenamefont {Conway}\ \textit {et~al.}(1995)\citenamefont {Conway},
  \citenamefont {Duda}, \citenamefont {Cheng}, \citenamefont {Hendrix},\ and\
  \citenamefont {Steven}}]{conway2}%
  \BibitemOpen
  \bibfield  {author} {\bibinfo {author} {\bibfnamefont {J.}~\bibnamefont
  {Conway}}, \bibinfo {author} {\bibfnamefont {R.}~\bibnamefont {Duda}},
  \bibinfo {author} {\bibfnamefont {N.}~\bibnamefont {Cheng}}, \bibinfo
  {author} {\bibfnamefont {R.}~\bibnamefont {Hendrix}}, \ and\ \bibinfo
  {author} {\bibfnamefont {A.}~\bibnamefont {Steven}},\ }\href@noop {}
  {\bibfield  {journal} {\bibinfo  {journal} {Journal of molecular biology}\
  }\textbf {\bibinfo {volume} {253}},\ \bibinfo {pages} {86} (\bibinfo {year}
  {1995})}\BibitemShut {NoStop}%
\bibitem [{\citenamefont {Steven}\ \textit {et~al.}(1997)\citenamefont {Steven},
  \citenamefont {Trus}, \citenamefont {Booy}, \citenamefont {Cheng},
  \citenamefont {Zlotnick}, \citenamefont {Caston},\ and\ \citenamefont
  {Conway}}]{Steven}%
  \BibitemOpen
  \bibfield  {author} {\bibinfo {author} {\bibfnamefont {A.~C.}\ \bibnamefont
  {Steven}}, \bibinfo {author} {\bibfnamefont {B.}~\bibnamefont {Trus}},
  \bibinfo {author} {\bibfnamefont {F.}~\bibnamefont {Booy}}, \bibinfo {author}
  {\bibfnamefont {N.}~\bibnamefont {Cheng}}, \bibinfo {author} {\bibfnamefont
  {A.}~\bibnamefont {Zlotnick}}, \bibinfo {author} {\bibfnamefont
  {J.}~\bibnamefont {Caston}}, \ and\ \bibinfo {author} {\bibfnamefont
  {J.}~\bibnamefont {Conway}},\ }\href@noop {} {\bibfield  {journal} {\bibinfo
  {journal} {The FASEB journal}\ }\textbf {\bibinfo {volume} {11}},\ \bibinfo
  {pages} {733} (\bibinfo {year} {1997})}\BibitemShut {NoStop}%
\bibitem [{\citenamefont {White}\ \textit {et~al.}(2012)\citenamefont {White},
  \citenamefont {Sherman}, \citenamefont {Brasiles}, \citenamefont {Jacquet},
  \citenamefont {Seavers}, \citenamefont {Tavares},\ and\ \citenamefont
  {Orlova}}]{White}%
  \BibitemOpen
  \bibfield  {author} {\bibinfo {author} {\bibfnamefont {H.}~\bibnamefont
  {White}}, \bibinfo {author} {\bibfnamefont {M.}~\bibnamefont {Sherman}},
  \bibinfo {author} {\bibfnamefont {S.}~\bibnamefont {Brasiles}}, \bibinfo
  {author} {\bibfnamefont {E.}~\bibnamefont {Jacquet}}, \bibinfo {author}
  {\bibfnamefont {P.}~\bibnamefont {Seavers}}, \bibinfo {author} {\bibfnamefont
  {P.}~\bibnamefont {Tavares}}, \ and\ \bibinfo {author} {\bibfnamefont
  {E.}~\bibnamefont {Orlova}},\ }\href@noop {} {\bibfield  {journal} {\bibinfo
  {journal} {J. Virol.}\ }\textbf {\bibinfo {volume} {86}},\ \bibinfo {pages}
  {6768} (\bibinfo {year} {2012})}\BibitemShut {NoStop}%
\bibitem [{Note7()}]{Note7}%
  \BibitemOpen
  \bibinfo {note} {Tresset, G et al. Phys. Rev. Applied, accepted for
  publication.}\BibitemShut {Stop}%
\bibitem [{\citenamefont {Lidmar}\ \textit {et~al.}(2003)\citenamefont {Lidmar},
  \citenamefont {Mirny},\ and\ \citenamefont {Nelson}}]{Lidmar}%
  \BibitemOpen
  \bibfield  {author} {\bibinfo {author} {\bibfnamefont {J.}~\bibnamefont
  {Lidmar}}, \bibinfo {author} {\bibfnamefont {L.}~\bibnamefont {Mirny}}, \
  and\ \bibinfo {author} {\bibfnamefont {D.}~\bibnamefont {Nelson}},\
  }\href@noop {} {\bibfield  {journal} {\bibinfo  {journal} {Physical Review
  E}\ }\textbf {\bibinfo {volume} {68}},\ \bibinfo {pages} {051910} (\bibinfo
  {year} {2003})}\BibitemShut {NoStop}%
\bibitem [{\citenamefont {Zandi}\ \textit {et~al.}(2004)\citenamefont {Zandi},
  \citenamefont {Reguera}, \citenamefont {Bruinsma}, \citenamefont {Gelbart},\
  and\ \citenamefont {Rudnick}}]{Zandi}%
  \BibitemOpen
  \bibfield  {author} {\bibinfo {author} {\bibfnamefont {R.}~\bibnamefont
  {Zandi}}, \bibinfo {author} {\bibfnamefont {D.}~\bibnamefont {Reguera}},
  \bibinfo {author} {\bibfnamefont {R.}~\bibnamefont {Bruinsma}}, \bibinfo
  {author} {\bibfnamefont {W.}~\bibnamefont {Gelbart}}, \ and\ \bibinfo
  {author} {\bibfnamefont {J.}~\bibnamefont {Rudnick}},\ }\href@noop {}
  {\bibfield  {journal} {\bibinfo  {journal} {Proc Natl Acad Sci U S A.}\
  }\textbf {\bibinfo {volume} {101}},\ \bibinfo {pages} {15556} (\bibinfo
  {year} {2004})}\BibitemShut {NoStop}%
\bibitem [{\citenamefont {Elrad}\ and\ \citenamefont {Hagan}(2010)}]{Hagan}%
  \BibitemOpen
  \bibfield  {author} {\bibinfo {author} {\bibfnamefont {O.}~\bibnamefont
  {Elrad}}\ and\ \bibinfo {author} {\bibfnamefont {M.}~\bibnamefont {Hagan}},\
  }\href@noop {} {\bibfield  {journal} {\bibinfo  {journal} {Physical Biology}\
  }\textbf {\bibinfo {volume} {7}},\ \bibinfo {pages} {045003} (\bibinfo {year}
  {2010})}\BibitemShut {NoStop}%
\bibitem [{\citenamefont {Landau}(1937)}]{Landau1}%
  \BibitemOpen
  \bibfield  {author} {\bibinfo {author} {\bibfnamefont {M.}~\bibnamefont
  {Landau}},\ }\href@noop {} {\bibfield  {journal} {\bibinfo  {journal} {Zh.
  Eksp. Teor. Fiz.}\ }\textbf {\bibinfo {volume} {7}},\ \bibinfo {pages} {19}
  (\bibinfo {year} {1937})}\BibitemShut {NoStop}%
\bibitem [{Note8()}]{Note8}%
  \BibitemOpen
  \bibinfo {note} {Translated and reprinted in Landau L.D. Collected Papers
  (Nauka, Moscow, 1969), Vol. 1, pp. 234–252. Amusingly, Landau compares in
  this paper different irreps to different \protect \textit {races}. In this
  context, the proposed theory could be viewed as a demonstration of the
  benefits of increased ethnic diversity.}\BibitemShut {Stop}%
\bibitem [{\citenamefont {Landau}\ and\ \citenamefont
  {Lifshitz}(1980)}]{Landau2}%
  \BibitemOpen
  \bibfield  {author} {\bibinfo {author} {\bibfnamefont {L.~D.}\ \bibnamefont
  {Landau}}\ and\ \bibinfo {author} {\bibfnamefont {E.~M.}\ \bibnamefont
  {Lifshitz}},\ }\href@noop {} {\textit {\bibinfo {title} {Statistical physics,
  part I}}}\ (\bibinfo  {publisher} {Pergamon Press},\ \bibinfo {year}
  {1980})\BibitemShut {NoStop}%
\bibitem [{\citenamefont {Hamermesh}(1989)}]{Hamermesh}%
  \BibitemOpen
  \bibfield  {author} {\bibinfo {author} {\bibfnamefont {M.}~\bibnamefont
  {Hamermesh}},\ }\href {Publisher description
  http://www.loc.gov/catdir/description/dover032/89023257.html} {\textit
  {\bibinfo {title} {Group theory and its application to physical problems}}},\
  Dover books on physics and chemistry\ (\bibinfo  {publisher} {Dover
  Publications},\ \bibinfo {address} {New York},\ \bibinfo {year} {1989})\ pp.\
  \bibinfo {pages} {xv, 509 p.}\BibitemShut {Stop}%
\bibitem [{\citenamefont {Kingston}\ \textit {et~al.}(2001)\citenamefont
  {Kingston}, \citenamefont {Olson},\ and\ \citenamefont {Vogt}}]{Kingston}%
  \BibitemOpen
  \bibfield  {author} {\bibinfo {author} {\bibfnamefont {R.}~\bibnamefont
  {Kingston}}, \bibinfo {author} {\bibfnamefont {N.}~\bibnamefont {Olson}}, \
  and\ \bibinfo {author} {\bibfnamefont {V.}~\bibnamefont {Vogt}},\ }\href@noop
  {} {\bibfield  {journal} {\bibinfo  {journal} {J. Struct. Biol.}\ }\textbf
  {\bibinfo {volume} {136}},\ \bibinfo {pages} {67} (\bibinfo {year}
  {2001})}\BibitemShut {NoStop}%
\bibitem [{\citenamefont {Overby}\ \textit {et~al.}(2008)\citenamefont {Overby},
  \citenamefont {Pettersson}, \citenamefont {Grunewald},\ and\ \citenamefont
  {Huiskonen}}]{Overby}%
  \BibitemOpen
  \bibfield  {author} {\bibinfo {author} {\bibfnamefont {A.}~\bibnamefont
  {Overby}}, \bibinfo {author} {\bibfnamefont {R.}~\bibnamefont {Pettersson}},
  \bibinfo {author} {\bibfnamefont {K.}~\bibnamefont {Grunewald}}, \ and\
  \bibinfo {author} {\bibfnamefont {J.}~\bibnamefont {Huiskonen}},\ }\href@noop
  {} {\bibfield  {journal} {\bibinfo  {journal} {PNAS}\ }\textbf {\bibinfo
  {volume} {105}},\ \bibinfo {pages} {2375Ð2379} (\bibinfo {year}
  {2008})}\BibitemShut {NoStop}%
\end{thebibliography}%
\end{document}